\documentclass[sigplan,10pt]{acmart}

\acmYear{2018}
\acmISBN{} 
\acmDOI{} 

\setcopyright{none}

\usepackage{booktabs}   

\usepackage{amsmath,
            amsfonts,
            amssymb,
            latexsym,
            setspace,
            helvet}
\usepackage{extarrows}
\usepackage{amssymb}
\usepackage{graphicx}
\usepackage{color}
\usepackage[latin1]{inputenc}
\usepackage{subfigure}
\usepackage{extarrows}
\usepackage{multirow}
\usepackage{listings}
\usepackage{array}
\usepackage{multirow}

\usepackage{enumerate}

\usepackage{tikz}
\usepackage{verbatim}
\usepackage{pgfplots}
\usepackage{url}
\urldef{\mailsa}\path|{alfred.hofmann, ursula.barth, ingrid.haas, frank.holzwarth,|
\urldef{\mailsb}\path|anna.kramer, leonie.kunz, christine.reiss, nicole.sator,|
\urldef{\mailsc}\path|erika.siebert-cole, peter.strasser, lncs}@springer.com|

\newcommand{\forget}[1]{}

\newcommand{\lcolorbox}[2]{%
\hspace*{-\fboxsep}\colorbox{#1}{#2}%
}

\newcolumntype{"}{@{\hskip\tabcolsep\vrule width 2pt\hskip\tabcolsep}}

\def\hlinew#1{%
  \noalign{\ifnum0=`}\fi\hrule \@height #1 \futurelet
   \reserved@a\@xhline}

\begin{document}

\newtheorem{Def}{Definition}
\newtheorem{Expl}{Example}
\newtheorem{Thm}{Theorem}
\newtheorem{Lem}[Thm]{Lemma}
\newtheorem{fac}[Thm]{Fact}
\newtheorem{Cor}[Thm]{Corollary}

\def\squareforqed{\hbox{\rlap{$\sqcap$}$\sqcup$}}
\def\qed{\ifmmode\squareforqed\else{\unskip\nobreak\hfil
\penalty50\hskip1em\null\nobreak\hfil\squareforqed
\parfillskip=0pt\finalhyphendemerits=0\endgraf}\fi}

\newcounter{statement}
\def\stmnum{\hbox to .01pt{}\rlap{\rm \hskip -\displaywidth\thestatement.}}
\def\stm{\refstepcounter{statement}\topsep 2pt \trivlist \item[]\leavevmode
\hbox to\linewidth\bgroup $ \displaystyle \hskip\leftmargini}
\def\endstm{$\hfil \displaywidth\linewidth\stmnum\egroup \endtrivlist}

\newenvironment{Proof}{%
\begin{list}{}{\setlength{\topsep}{\jot}\setlength{\parsep}{\topsep}%
\addtolength{\parsep}{-0.3\parsep}\setlength{\leftmargin}{0pt}}%
\parindent 4ex
\item[]\setcounter{statement}{0}\textbf{Proof:}}{\end{list}}

\newcommand{\scarrow}[2]
      { \mathrel{\setbox0 \hbox{ ${\scriptstyle #1}$ }\displaystyle
            \mathop{\hbox to \wd0{$\Longrightarrow$}}\limits_{#1}^{#2}}\,\,\,
      }
\newcommand{\lscarrow}[2]
      { \mathrel{\setbox0 \hbox{ ${\scriptstyle #1}$ }\displaystyle
                 \mathop{\hbox to \wd0{$=\!=\!\Longrightarrow$}}\limits_{#1}^{#2}}
      }
\newcounter{LLN}
\newcommand{\beginit} {
                         \begin{list}{\hfill{\rm \arabic{LLN}.} }{\usecounter{LLN}
                         \setlength{\leftmargin}{\leftmarginii}
                         \setlength{\itemsep}{\smallskipamount}}}

\newbox\mystrutbox
\setbox\mystrutbox=\hbox{\vrule height10pt depth4pt width0pt}
\def\mystrut{\relax\ifmmode\copy\mystrutbox\else\unhcopy\mystrutbox\fi}
\newcommand{\lprf}[3]
       {\llap{{$#3$}\thinspace}{{\mystrut\displaystyle #1}
          \over
         {\mystrut\displaystyle #2}}}
\newcommand{\rprf}[3]
       {{{\mystrut\displaystyle #1}
          \over
         {\mystrut\displaystyle #2}}{\rlap{{$#3$}\thinspace}}}

\newcommand{\dotminus}{-\!\!\!\!^{\textstyle.}\!\!\!\!-}
\newcommand{\lit}[1]{{\rm [#1]}}
\newcommand{\henv}[1]{#1}
\newcommand{\setof}[2]{\{ #1 \: | \: #2 \}}
\newcommand{\bigsetof}[2]{\bigl\{ #1 \: \bigm| \: #2 \bigr\}}
\newcommand{\arrow}[1]{\stackrel{#1}{\longrightarrow}}
\newcommand{\sarrow}[1]{\stackrel{#1}{\Longrightarrow}}
\newcommand{\dobarrow}[1]{\stackrel{#1}{\Longrightarrow}}
\newcommand{\proc}{Pr}
\newcommand{\af}[1]{\forall{\rm F\;}#1}
\newcommand{\ef}[1]{\exists{\rm F\;}#1}
\newcommand{\ag}[1]{\forall{\rm G\;}#1}
\newcommand{\eg}[1]{\exists{\rm G\;}#1}
\newcommand{\varu}[3]{U^{#1}_{#2,#3}}
\newcommand{\varusfg}{\varu{s}{F}{G}}
\newcommand{\varuwfg}{\varu{w}{F}{G}}
\newcommand{\semu}[3]{{\cal U}^{#1}_{#2,#3}}
\newcommand{\semuwfg}{\semu{w}{F}{G}}
\newcommand{\semusfg}{\semu{s}{F}{G}}
\newcommand{\transu}[3]{{\cal T}^{#1}_{#2,#3}}
\newcommand{\transusfg}{\transu{s}{F}{G}}
\newcommand{\transuwfg}{\transu{w}{F}{G}}
\newcommand{\comp}[1]{{\cal C}(#1)}
\newcommand{\prc}[1]{{\cal P}(#1)}
\newcommand{\comppred}[1]{\prec_{#1}}
\newcommand{\bisim}[1]{\sim_{#1}}
\newcommand{\fat}[1]{\mbox{\bf #1}}
\newcommand{\tr}{\mbox{\rm tt}}
\newcommand{\fa}{\mbox{\rm ff}}
\newcommand{\act}{Act}
\newcommand{\impliess}{\: \Rightarrow \:}
\newcommand{\implied}{\Leftarrow}
\newcommand{\implieds}{\:\Leftarrow\:}
\newcommand{\iffs}{\:\Leftrightarrow\:}
\newcommand{\hviss}{\Leftrightarrow}
\newcommand{\sigent}{\sigma_{\entail{}}}
\newcommand{\ua}[2]{#1 \subseteq \sem{\D}#1 \cup #2}
\newcommand{\id}{\mbox{\rm Id}}
\newcommand{\powerset}[1]{{\Large \wp} (#1)}
\newcommand{\sem}[1]{\lbrack\!\lbrack #1 \rbrack\!\rbrack}
\newcommand{\abs}[1]{|\!| #1 |\!|}
\newcommand{\may}[1]{\langle #1 \rangle}
\newcommand{\wmay}[1]{\langle\!\langle #1 \rangle\!\rangle}
\newcommand{\wmust}[1]{\sem{#1}}
\newcommand{\until}[1]{[ #1 \rangle}
\newcommand{\smay}[1]{\langle\!\cdot #1 \cdot\!\rangle}
\newcommand{\must}[1]{[ #1 ]}
\newcommand{\smust}[1]{\lbrack\!\cdot #1 \cdot\!\rbrack}
\newcommand{\sat}[1]{\models_{#1}}
\newcommand{\mx}{{\rm max}}
\newcommand{\mn}{{\rm min}}
\newcommand{\satmn}{\sat{\mn}}
\newcommand{\satmx}{\sat{\mx}}
\newcommand{\sigmax}{\sigma_{\mx}}
\newcommand{\sigmin}{\sigma_{\mn}}
\newcommand{\entail}[1]{\vdash_{#1}}
\newcommand{\entmn}{\entail{\mn}}
\newcommand{\entmx}{\vdash}
\newcommand{\satt}[1]{|\!\!\!\equiv_{#1}}
\newcommand{\sattmx}{\satt{\mx}}
\newcommand{\sattmn}{\satt{\mn}}
\newcommand{\Dtu}{\D^{\tau}}
\newcommand{\Dtd}{\D_{\tau}}
\newcommand{\da}[2]{\semd #1 \cap #2 \subseteq #1}
\newcommand{\ass}[2]{#1 : #2}
\newcommand{\MM}[1]{{\cal M}_{#1}}
\newcommand{\D}{{\cal D}}
\newcommand{\mmid}{{\cal M}_{{\rm Id}}}
\newcommand{\og}{\wedge}
\newcommand{\eller}{\vee}
\newcommand{\semd}{\sem{\D}}
\newcommand{\M}{{\cal M}}
\newcommand{\infrule}[3]
           {\parbox{2cm}{ $$ {\frac {#1}{#2}}\hspace{.5cm}{#3} \hfill $$}}
\newcommand{\infrulegen}[4]
           {{#1}\hspace{.5cm}{\frac {#2}{#3}}\hspace{.5cm}{#4}}
\newcommand{\ent}{\entail{}}
\newcommand{\Gammah}{\widehat{\Gamma}}
\newcommand{\inv}[1]{\textsf{\rm Inv}(#1)}
\newcommand{\pos}[1]{\mbox{\rm Pos}(#1)}
\newcommand{\inva}{\inv{\may{a}\tr}}
\newcommand{\posa}{\pos{\must{a}\fa}}
\newcommand{\op}{{\cal O}}
\newcommand{\even}[1]{\mbox{\rm Even}(#1)}
\newcommand{\unic}{\bigr( \lbr\;|\;\hole\bsl p\bigr) \bsl\coin,\cof}
\newcommand{\live}{\mbox{\rm Live}}
\newcommand{\con}{\mbox{\rm Con}}
\newcommand{\com}{\mbox{\rm Com}}
\newcommand{\scom}{\mbox{{\cal Sc}}}
\newcommand{\dead}{\mbox{\rm Dead}}
\newcommand{\diver}{\mbox{\rm Div}}
\newcommand{\sUntil}[2]{\mbox{\rm Unt}^s(#1,#2)}
\newcommand{\wUntil}[2]{\mbox{\rm Unt}^w(#1,#2)}
\newcommand{\saf}[1]{\mbox{\rm Saf}(#1)}
\newcommand{\uni}{\mbox{\sl Uni}}
\newcommand{\staff}{\mbox{\sl Staff}}
\newcommand{\equip}{\mbox{\sl Equip}}
\newcommand{\acc}{\mbox{acc}}
\newcommand{\del}{\mbox{del}}
\newcommand{\wip}[2]{\mbox{\rm wip}(#1,#2)}
\newcommand{\sop}[2]{\mbox{\rm sop}(#1,#2)}
\newcommand{\cuni}{C_{\mbox{\rm uni}}}
\newcommand{\lbr}{\mbox{\rm lbr}}
\newcommand{\carrow}[2]{\raisebox{0.2ex}{$\,\,\,\,\,{#2\atop #1}
\!\!\!\!\!\!\!\!\arrow{}\,\,
$}}
\newcommand{\ccarrow}[2]{\raisebox{0.2ex}{$\,\,\,\,\,\,{#2\atop #1}
\!\!\!\!\!\!\!\!\!\arrow{}\!\!\!\!\!\!\!\!\arrow{}\,\,
$}}
\newcommand{\dccarrow}[2]{\raisebox{0.2ex}{$\,\,\,\,\,\,{#2\atop #1}
\!\!\!\!\!\!\!\!\!\arrow{}\!\!\!\!\!\!\!\!\arrow{}_\C\,\,
$}}
\newcommand{\cdarrow}[2]{\raisebox{0.2ex}{$\,\,\,\,\,\,{#2\atop #1}
\!\!\!\!\!\!\!\!\!\arrow{}_\C\,
$}}
\newcommand{\dcarrow}[2]{{\raisebox{-1.2ex}{
                         $\stackrel{#2}{\stackrel{\longrightarrow_\Diamond}
                          {\scriptstyle #1}}$  }}}
\newcommand{\rearrow}[2]{\raisebox{-1.2ex}{
                         $\stackrel{#1}{\stackrel{\longrightarrow}
                          {\scriptstyle #2}}$  }}
\newcommand{\bsl}{\backslash}
\newcommand{\hole}{\mbox{$[\;]$}}
\newcommand{\coin}{\mbox{coin}}
\newcommand{\cof}{\mbox{cof}}
\newcommand{\unicp}{\bigr( \cof .(\lbr + p.\lbr) \;|\;\hole\bsl p\bigr)
                                   \bsl \coin,\cof}
\newcommand{\unicpp}{\bigr( (\lbr + p.\lbr)\;|\;\hole\bsl p\bigr)
                             \bsl \coin,\cof}
\newcommand{\synen}[2]{#1\,\,{\bf{\sf with}}\,\,#2}
\newcommand{\emax}[2]{\mbox{{\bf {\sf max }}}#1\,\,{\bf{\sf with}}\,\,#2}
\newcommand{\emin}[2]{\mbox{{\bf {\sf min }}}#1\,\,{\bf{\sf with}}\,\,#2}
\newcommand{\lmax}{\mbox{{\sf max}\,}}
\newcommand{\lmin}{\mbox{{\sf min}\,}}
\newcommand{\Dsem}[1]{{\sf D} \lbrack\!\lbrack #1 \rbrack\!\rbrack}
\newcommand{\Delsem}[1]{{\sf \Delta} \lbrack\!\lbrack #1 \rbrack\!\rbrack}
\newcommand{\Asem}[1]{{\sf E} \lbrack\!\lbrack #1 \rbrack\!\rbrack}
\newcommand{\Lsem}[1]{{\sf L} \lbrack\!\lbrack #1 \rbrack\!\rbrack}
\newcommand{\Nsem}[1]{{\sf N} \lbrack\!\lbrack #1 \rbrack\!\rbrack}
\newcommand{\DMTSt}{\mbox{DMTS$^2$}\,}
\newcommand{\DMTSs}{\mbox{DMTS$^\ast$}\,}
\newcommand{\conf}[2]{\langle #1,#2\rangle}
\newcommand{\chop}[1]{\mbox{{\tt chop}$(#1)$}}
\newcommand{\tl}[1]{\mbox{{\tt tl}$(#1)$}}
\newcommand{\eemax}[1]{\mbox{{\bf {\sf max }}}#1}
\newcommand{\eemin}[1]{\mbox{{\bf {\sf min }}}#1}
\newcommand{\pre}{\mbox{\,{\tt pre}\,}}
\newcommand{\conft}[1]{\langle #1\rangle}
\newcommand{\incon}{IC}
\newcommand{\hml}{{\cal M}}
\newcommand{\depth}{depth}
\newcommand{\dsat}{{\,\mbox{{\tt sat}}}}
\newcommand{\branb}{{\approx_b}}
\newcommand{\V}{{\cal V}}
\newcommand{\Z}{{\cal Z}}
\newcommand{\ins}[3]{\vdash^?_{#3}#1\colon#2}
\newcommand{\insp}[3]{\vdash^{#3}#1\colon#2}
\newcommand{\insn}[3]{\not\vdash_{#3}#1\colon#2}
\newcommand{\wi}[2]{{\cal W}(#1,#2)}


\newcommand{\DEF}               {\stackrel{\rm def}{=}}
\newcommand{\invo}              {\textsf{inv}}
\newcommand{\res}               {\mathit{res}}
\newcommand{\lin}               {\textsf{lin}}
\newcommand{\Abs}               {\textsf{Abs}}
\newcommand{\client}            {\textsf{Client}}
\newcommand{\Sys}               {\textsf{Sys}}
\newcommand{\Systau}            {\textsf{Sys1}}
\newcommand{\Sysinv}            {\textsf{Sys2}}
\newcommand{\Sysres}            {\textsf{Sys3}}
\newcommand{\Sysobj}            {\textsf{Sys4}}
\newcommand{\emp}               {\varepsilon}
\newcommand{\Ex}                {\bf Ex}
\newcommand{\Eva}               {\bf Eva}
\newcommand{\true}              {\mathit{true}}
\newcommand{\false}             {\mathit{false}}

\newcommand{\StackOp}           {\mathit{StackOp}}
\newcommand{\LesOp}             {\mathit{LesOp}}
\newcommand{\TryStackOp}        {\mathit{TryStackOp}}
\newcommand{\TryCollision}      {\mathit{TryCollision}}
\newcommand\ignore[1]{}
\newcommand{\call}              {\texttt{call}}
\newcommand{\ret}               {\texttt{ret}}

\newcommand{\xhookrightarrow}[1]  {\stackrel{#1}{\hookrightarrow}}


\definecolor{ablue}{RGB}{0,153,225}
\definecolor{aorange}{RGB}{225,64,0}
\definecolor{apurple}{RGB}{163,143,196}
\definecolor{corg}{RGB}{204,102,0}

\newcommand{\xx}[1]{\textcolor{ablue}{\bf{#1}}}
\newcommand{\jp}[1]{\textcolor{aorange}{\bf{#1}}}
\newcommand{\st}[1]{\textcolor{apurple}{\bf{#1}}}

\newcommand{\atl}[1]{\texttt{\small{Line #1}}}
\newcommand{\atls}[1]{\texttt{\small{Lines #1}}}

\newenvironment{narrow}[2]{%
\begin{list}{}{%
\setlength{\topsep}{0pt}%
\setlength{\leftmargin}{#1}%
\setlength{\rightmargin}{#2}%
\setlength{\listparindent}{\parindent}%
\setlength{\itemindent}{\parindent}%
\setlength{\parsep}{\parskip}}%
\item[]}{\end{list}}


\title[]{The Effect Race in Fine-Grained Concurrency}         



\author{Xiaoxiao Yang}
\orcid{nnnn-nnnn-nnnn-nnnn}             
\affiliation{
  \institution{Institute of Software, Chinese Academy of Sciences}            
  \city{Beijing}
 \country{China}                    
}
\email{xxyang@ios.ac.cn}          

\begin{abstract}
  Most existed work require knowledge about the effect of program instructions (or statements) to analyze and verify algorithms. In this paper, by revealing some findings on executions of object programs, we define two basic concepts -- effect equivalence relation and effect race relation. Further, we show three effect theorems about the race and histories. The core result is that the effect race relation is the accurate relation to capture the internal steps, of which precedence orders are the reason to cause chaotic histories. In addition,  the concept -- linearization points -- widely used in the object verification, is defined formally as the typical effect race relation. These results provide a clear basis for analyzing intricate fine-grained executions. We conduct a lot of experiments on real object algorithms to show the accuracy and efficiency of these definitions in practice. A simple quantitative analysis method for these algorithms is also proposed.
\end{abstract}

\keywords{Concurrent data structures, Branching Bisimulation, Verification, Effects, Linearizability}  

\maketitle

\section{Introduction}








\subsection{Overview}

Most highly-optimized concurrent data structures (also called concurrent objects) are designed by using fine-grained synchronization techniques (e.g., CAS, coupling-locks), which involve intricate interleavings.
The main correctness conditions of concurrent objects, e.g., sequential consistency \cite{Lamport82} and linearizability \cite{Herlihy90}, are defined on the coarse-grained notion of the \emph{history} -- a \emph{finite} sequence of call and return actions.
These actions as the interactions with clients are called \textit{visible actions},   and all the internal program instructions are regarded as \emph{silence actions}.
We found that these visible actions on the history have the following main features:

\begin{enumerate}[(1)]
  \item visible actions are acquired from executions.
  \item no visible actions get access to the shared state.
\end{enumerate}

When a shared mutable state is modified by an internal instruction,  the term ``effect'' is utilized by programmers and verifiers to express the impact of the state change on the outside world. Due to the race condition, the shared states in the fine-grained program can be modified in many different execution orders. This brings a lot of disordered and unexpected histories. Algorithm analysis starts from the histories.
How to correctly understand the effect of state changes on visible actions is the crux for the verification.
For example,
linearization points (LPs) \cite{Herlihy08} are the typical instructions to represent the effect of method calls and have being used as the main means to prove the linearizability (e.g., \cite{Victor08,Schellhorn12}).
However, locating LPs is a bottleneck when conducing a proof.
The existed work  can only give the informal descriptions of effects of LPs (e.g., future-dependent LPs or helping~\cite{Feng13}) in terms of the observed phenomenon on individual algorithms.


In this paper, we provide some formal basis for analyzing the effect of concurrent programs.
By reflection on LPs, we first reveal that the ordinary trace equivalence cannot precisely perceive the effect of LPs, and  \emph{branching potentials} play a vital role to determine effects. This motivates us to define the effect equivalence relation on states based on the max-trace equivalence~\cite{Glabbeek96}.

The core definition in the work is the effect race relation, a binary relation on internal steps.
Fine-grained concurrency involves a lot of race that access the shared state.
Effect theorems in the paper reveal that the effect race relation is the accurate relation to capture the internal steps, of which different execution orders are the reason to cause disorder histories. This finding provides a clear clue to analyze complex fine-grained algorithms.

We also find and define a neat structure, called effect structure, which is a subset of executions of object programs. Effect structures
establish a connection between the effect of each non-stutter step and executions. It is shown that each step relating the states that are not effect equivalent is critical and represents a race with another non-stutter step in the effect structure.
The existence of effect structures assures the results of the Effect theorems.

It needs to be emphasized that these results have no concern with the implementation details, but simply rely on two conditions: (i) visible actions in a system satisfy the above features (1) and (2), and (ii) the stuttering property guaranteed by the effect equivalence relation. So these results are suitable for the general algorithm.


We further formalize LPs by means of the effect race relation, which allows understanding LPs more clearly.

Since the max-trace equivalence is equivalent to the branching bisimilar \cite{Glabbeek96}, in practice, the effect equivalence relation can be computed efficiently by the branching bisimulation equivalence. We have conducted experiments on various well-known concurrent data structures. Experimental results validate the accuracy and efficiency of these definitions on analyzing real algorithms.


\subsection{Trace equivalence and effect equivalence}

The subtlety of fine-grained executions on the effect can be illustrated using
the classic Herlihy and Wing queue algorithm \cite{Herlihy90},
shown in Figure \ref{pic-6}.
The queue's representation is an indexed array $\mathtt{AR}$ with $\mathtt{back}$ as the index to denote
the next unused slot in $\mathtt{AR}$. Each slot is initialized to a value $\mathtt{null}$, and
$\mathtt{back}$ is initialized to 1. The queue has two methods, $\mathtt{Enq}$ and $\mathtt{Deq}$.
An $\mathtt{Enq}$ execution contains two steps: first gets a copy of $\mathtt{back}$ and increases $\mathtt{back}$;
then stores an element at $\mathtt{AR[i]}$.
A $\mathtt{Deq}$ execution visits $\mathtt{AR}$ in ascending order,
starting from index $1$ and ending at $\mathtt{back-1}$.
If $\mathtt{Deq}$ finds a non-$\mathtt{null}$ value at slot $\mathtt{i}$, it will return the value of $\mathtt{AR[i]}$, otherwise it tries the next slot. If no element is found, then $\mathtt{Deq}$ will restart. Each execution step of a method call is atomic and is interleaved with steps of other concurrent method calls.
\begin{figure}[htpb]
\begin{minipage}{.70\textwidth}
\vspace{-2ex}
\begin{lstlisting}[basicstyle=\scriptsize\ttfamily]
E0 Enq(x:T) {
E1 (i, back):=(back, back+1);  /* increment */
E2 AR[i]:=x; /* store */
E3 return
E4 }
\end{lstlisting}
\begin{lstlisting}[basicstyle=\scriptsize\ttfamily]
D0 Deq() {
D1  while true do {
D2    range := back;
D3    for (0 < i < range) do {
D4      (x, AR[i]):=(AR[i], null);  /* swap */
D5      if (x != null) then return (x)
D6 } } }
\end{lstlisting}
\end{minipage}
\vspace{-2ex}
\caption{Herlihy and Wing queue.} \label{pic-6} 
\end{figure}

\begin{figure*}[htpb]
  \centering
  \vspace{-1.8ex}
  \includegraphics[scale=.46]{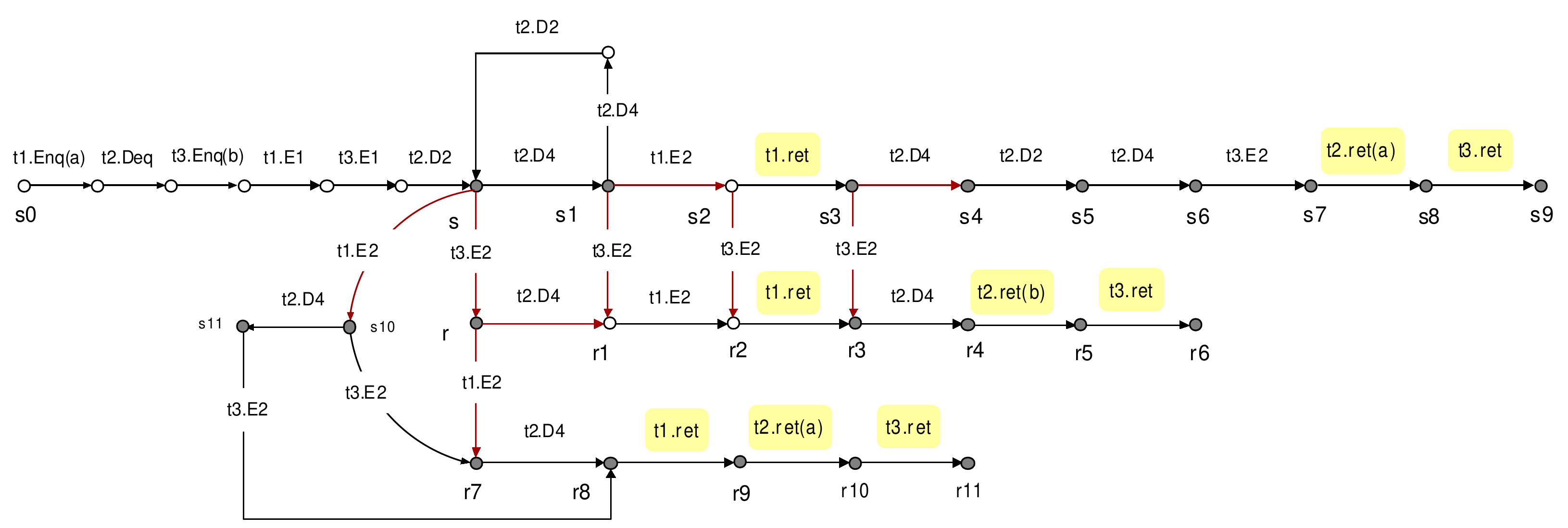}\\
  \caption{A part of the transition system for the Herlihy and Wing queue.}\label{HW-traces}
\end{figure*}

The behaviour of the concurrent system can be modeled as a labeled transition system.
It is common to understand an algorithm by observing possible (finite) executions at each state (e.g., \cite{DBLP:journals/tocl/SchellhornDW14}).
For the Herlihy and Wing queue example, consider a system of three threads $\mathtt{t_1}$, $\mathtt{t_2}$ and $\mathtt{t_3}$,
with $\mathtt{t_1}$ executing $\mathtt{Enq(a)}$, $\mathtt{t_2}$ executing $\mathtt{Deq}$ and $\mathtt{t_3}$ executing $\mathtt{Enq(b)}$ concurrently.
A part of the transition graph generated from the system is depicted in Figure~\ref{HW-traces},
where $s_0$ is the initial state, and each step on executions is labeled with the corresponding instructions (i.e., $\mathtt{Ei}$ or $\mathtt{Di}$).
The call and return actions of the $\mathtt{Enq}$ method (i.e., $\mathtt{E0}$ and $\mathtt{E3}$),
of a thread $\mathtt{t}$ are denoted by $\mathtt{t.Enq(v)}$ and $\mathtt{t.ret}$ respectively (similar notions for the $\mathtt{Deq}$).
All internal computation steps are invisible, and denoted by $\tau$.
The states marked with $\circ$ have some additional $\tau$ transitions which
are irrelevant to the discussions and hence omitted.

An interesting step is  ${\small s \xlongrightarrow{} r}$ with instruction $\tau(\mathtt{t_3.E_2})$. It is the LP for method call $\mathtt{t_3.Enq(b)}$, and takes effect to change the empty queue by storing $\mathtt {b}$ at $\mathtt {AR[2]}$ successfully.
The effect of the LP is witnessed by the return action $\mathtt{t_2.ret(b)}$ on ${\small r_4 \xlongrightarrow{} r_5}$ transition.
However, for the LP ${\small s \xlongrightarrow{} r}$, traces cannot distinguish the effects of $s$ and $r$.
By omitting states and $\tau$ transitions on executions, it is not difficult to see that $s$ and $r$ have the same set of traces.
We use $\xLongrightarrow{\tau}$ to denote a sequence of $\tau$ transitions.
First, every trace of $r$ is a trace of $s$. The other direction of inclusion can be seen by the following executions from $s$, that is, the trace of $s$ below

\begin{quote}
${\small s \xLongrightarrow{\tau} s_2 \xlongrightarrow{\mathtt{t_1.ret}} s_3 \xLongrightarrow{\tau} s_7 \xlongrightarrow{\mathtt{t_2.ret(a)}} s_8  \xlongrightarrow{\mathtt{t_3.ret}} s_9}$ and \\
${\small s \xLongrightarrow{\tau} s_2 \xlongrightarrow{\mathtt{t_1.ret}} s_3 \xLongrightarrow{\tau} r_4 \xlongrightarrow{\mathtt{t_2.ret(b)}} r_5  \xlongrightarrow{\mathtt{t_3.ret}} r_6}$
\end{quote}
can be matched, by the following traces from $r$
\begin{quote}
${\small r \xLongrightarrow{\tau} r_2 \xlongrightarrow{\mathtt{t_1.ret}} r_3 \xlongrightarrow{\tau} r_4 \xlongrightarrow{\mathtt{t_2.ret(b)}} r_5  \xlongrightarrow{\mathtt{t_3.ret}} r_6}$ and \\
${\small r \xLongrightarrow{\tau} r_8 \xlongrightarrow{\mathtt{t_1.ret}} r_9  \xlongrightarrow{\mathtt{t_2.ret(a)}} r_{10}  \xlongrightarrow{\mathtt{t_3.ret}} r_{11}}$
\end{quote}

This is a well-known phenomenon in concurrency: $s$ and $r$ have the same trace set, but after $\mathtt{t_1.ret}$, the trace set of $s_3$
cannot be matched by any trace sets at $r_3$ or $r_9$. Thus, different effects of $s$ and $r$ are captured by branches,
that is, the trace set of $s_3$, $r_3$ and $r_9$ on the subsequent executions from $s$ and $r$, respectively.
Therefore, branching potentials play a vital role in determining the effect of the fine-grained implementation. This inspires us to characterize the effect equivalence relation based on branching potentials.
\\
\noindent
\emph{\textbf{Organizations.}}
Section 2 briefly reviews object systems and histories. Section 3 defines the effect equivalence relation.
Section 4 presents the effect race relation and effect structures. Section 5 provides three effect theorems. Section 6 shows branching bisimulation. Section 7 analyzes the effect of real algorithms. Section 8 defines linearization points. Section 9 presents a simple quantitative analysis for algorithms. Section 10 concludes.

\section{Object Systems and Histories}

\subsection{Object Systems}

The behaviors of a concurrent object can be adequately described as a labeled transition system.
We assume there is a language for describing concurrent algorithms, and
the language is equipped with an operational semantics to generate
labeled transition systems as defined below. To generate an object's behaviour, we use \emph{the most general clients}  \cite{Gotsman11,Liu13}, which
repeatedly invoke an object's methods in any order and with any possible parameters.

In the context, ``object systems''  refer to either the transition systems
or the program texts. Let $\mathtt{m(n)}$ denote method $\mathtt{m}$ with parameter $\mathtt{n}$. For simplicity,
all methods will take one parameter and return an integer value.


\begin{definition}[Labeled transition systems for concurrent objects] \label{lts}
A \emph{labled transition system} $\Delta$ is a quadruple $(S, \longrightarrow, {\mathcal A}, s_0)$
where
\begin{itemize}
  \item[$\bullet$] $S$ is the set of states,
 \item[$\bullet$]
    ${\mathcal A} = \{(\mathtt{t},\mathtt{call},\mathtt{m(n))}, (\mathtt{t},\mathtt{ret(n')},\mathtt{m}), (\mathtt{t},\tau) \mid  \mathtt{t} \in \{1 \ldots k\}$, where $k$ is the number of threads\} is the set of actions.
  \item[$\bullet$] $\longrightarrow\ \subseteq S \times {\mathcal A} \times S$ is the transition relation,
   \item[$\bullet$] $s_0 \in S$ is the initial state.
\end{itemize}
\qed
\end{definition}
We shall write $s  \xrightarrow{a} s'$ to abbreviate $(s, a, s') \in \longrightarrow$.

When analysing the behaviours of a concurrent object, we are interested in
the interactions (i.e., call and return) between the object and its clients,
while the internal instructions of the object are considered invisible and modeled by silence action $\tau$.


We write $s  \xrightarrow{\tau} s'$ to mean $s  \xrightarrow{(t,\tau)} s'$
for some $t$.
A \emph{path} $\rho(s)$ starting at a state $s$ of an object system
is a finite or infinite sequence $s \xlongrightarrow{a_1} s_{1} \xlongrightarrow{a_2} s_{2} \xlongrightarrow{a_3} \cdots$.
An execution is a path starting from the initial state, which represents an entire computation
of the object system. A {\em trace} of state $s$ is a sequence of visible actions
obtained from a path of $s$ by omitting states and invisible actions.

\subsection{Histories}

A history is a finite execution traces consisting of call and return actions,
to model the behavior of concurrent objects.

A history is \emph{sequential} if
(1) it starts with a method call, (2) call actions and return actions alternate in the history, (3)
each return matches immediately the previous call.
A sequential history is \emph{legal} if it respects the sequential specification of the object.
If $H$ is a history and $t$ a thread, then the projection of $H$ on $t$, written $H|t$,
is called the subhistory of $H$ on $t$.
An operation $\mathtt{e}$ is a pair which consists of an invocation event $\mathtt{(t,call,m(n))}$ and the matching response event $\mathtt{(t,ret(n'),{m})}$.
We shall use $\mathtt{e.call}$ and $\mathtt{e.ret}$ to denote, respectively, the invocation and
response events of an operation $\mathtt{e}$.
The operation ordering in $H$ can be formally described using an irreflexive partial order $<_H$ by requiring that
$(\mathtt{e}, \mathtt{e'}) \in\ <_H$ if $\mathtt{e.ret}$ precedes $\mathtt{e'.call}$ in $H$.
Operations that are not related by $<_H$ are said to be \emph{concurrent} (or overlapping).
If $H$ is sequential then $<_{H}$ is a total order.

The key idea behind the correctness conditions of concurrent objects (e.g., linearizability) is to compare concurrent histories to legal sequential histories. We show the linearizability relation on histories~\cite{Herlihy90,pldi10}.

\begin{definition}[Linearizability relation on histories]
$H \sqsubseteq_{\textsf{lin}} S$, read ``$H$ is linearizable \emph{w.r.t.} $S$'', if (1) $S$ is sequential, (2) $H|t = S|t$ for each thread $t$, and (3) $<_H~ \subseteq ~ <_{S}$.
\qed
\end{definition}


For the sequential specification $\Gamma$ and object system $\Delta$,
we use ${\mathcal H}(\Gamma)$ and ${\mathcal H}(\Delta)$ to denote the set of all histories of $\Gamma$ and $\Delta$ respectively. An execution or history is completed if there is no pending call.  Let $H(\sigma)$ denote a completed history obtained from an execution $\sigma$.  A history $H(\sigma)$ is said to be equivalent to a legal sequential history $S$ iff ${\small H(\sigma) \sqsubseteq_{\textsf{lin}} S}$ and ${\small S \in {\mathcal H}(\Gamma)}$.

In next sections, due to the space limit, we only show proofs of some results.

\section{Effect Equivalence}

This section explains the motivation on the effect equivalence from the perspective of executions,
and then formalize it by using the max-trace equivalence, and finally define stutter steps in object systems.

\subsection{Motivation on the Effect Equivalence}

In the object implementation, whether a step \emph{takes effect} is the crux to form a completed history.
Therefore, if a step ${\small s \xlongrightarrow{} r}$ is stutter for the object execution, then $s$ and $r$ should have the same effect. 
The effect change of a step is related to the change of object states, which can be captured by observing visible actions in the system.
Intuitively, a step ${\small s \xlongrightarrow{} r}$ keeps the same effect implies that, for any path $\rho(s)$ from $s$,
there exists an path $\rho(r)$ from $r$ such that $\rho(r)$ can``match" $\rho(s)$.
Informally, the``match" implies that
\begin{enumerate}
  \item $s$ and $r$ are trace equivalent;
  \item $\rho(s)$ and $\rho(r)$ are effect stutter equivalent.
\end{enumerate}

The Herlihy and Wing queue example shows the importance of branching potentials for the effect of method calls.
We now show, to assure the stutter equivalence \emph{w.r.t.} the effect change, it is necessary to consider traces of
each intermediate state of executions.






\begin{Expl}\rm
In Figure~\ref{effect-stutter}, it is easy to see states $1$ and $2$ are trace equivalent. But ${\small 1 \xlongrightarrow{} 3}$ and ${\small 2 \xlongrightarrow{} 4 \xlongrightarrow{} 3}$ are not stutter equivalent w.r.t. the effect change, since the trace set of intermediate state $4$ is neither equivalent to the trace set of $1$, nor to the trace set of $3$.
Therefore, the path ${\small 2 \xlongrightarrow{} 4 \xlongrightarrow{} 3}$ cannot match ${\small 1 \xlongrightarrow{} 3}$.
\qed
\begin{figure}[htpb]
  \centering
  \includegraphics[width=12em]{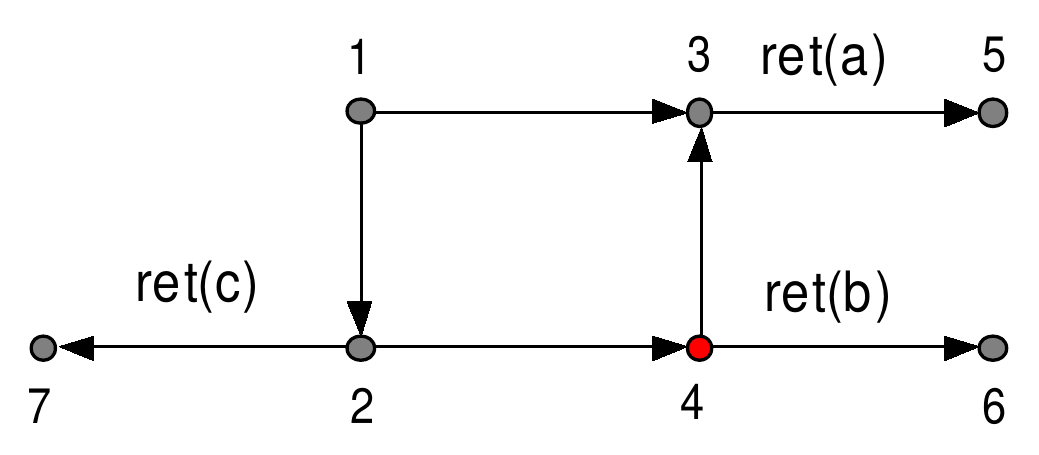}\\
  \caption{The effect of intermediate state $4$.}\label{effect-stutter}
\end{figure}
\end{Expl}

The above example shows that the effect equivalence relies on the traces of intermediate states over a path, and further,
for these intermediate states, their effects also depend on each intermediate state over the subsequent executions from that state.
Thus, the effect equivalence relation should be defined in an inductive way.
We show the intuitive idea as follows.

Let ${\small T^1(s)}$ denote the trace set of state $s$. If $s$ and $r$ are effect equivalent, then
(i) $s$ and $r$ are trace equivalent, i.e., ${\small T^1(s) = T^1(r)}$.
(ii) any path $\rho(s)$ from $s$ should be effect stutter equivalent to path $\rho(r)$ from $r$, and vice versa.
To guarantee it, we compare the effect equivalence of each intermediate state over $\rho(s)$ and $\rho(r)$. 

Let ${\scriptsize \rho(s) = s \xlongrightarrow{a_1} s_1 \xlongrightarrow{a_2} s_2 }$ ${\scriptsize \xlongrightarrow{a_3} \cdots \xlongrightarrow{a_n} s_n}$ and ${\scriptsize \rho(r) = r \xlongrightarrow{a_1}}$ ${\scriptsize r_1 }$ 
${\scriptsize \xlongrightarrow{a_2} r_2 \xlongrightarrow{a_3} \cdots \xlongrightarrow{a_n} r_m}$.
First, we compute the ordinary traces of each $s_i$ and $r_i$, and form sequence ${\scriptsize \rho_1 = (T^1(s), a_1, T^1(s_1), \cdots)}$ obtained from $\rho(s)$, and
${\scriptsize \rho_2 = (T^1(r), a_1, T^1(r_1),}$ ${\scriptsize \cdots)}$ obtained from $\rho(r)$. Thus,  ${\scriptsize \rho_1}$ is required to be stutter equivalent to ${\scriptsize \rho_2}$. Let ${\small T^2(s)}$ (resp.${\small T^2(r)}$) denote ${\scriptsize \rho_1}$ (resp.${\small \rho_2}$).
The effect equivalence of $s_i$ and $r_i$ will influence the effect equivalence of starting states $s$ and $r$.
To guarantee the effect equivalence of ${\small s_1,s_2, r_1,r_2,\cdots}$, we not only see the ordinary trace sets of ${\small s_1,s_2, r_1,r_2,\cdots}$,
but also the trace equivalence of the intermediate states over paths from these states on $\rho(s)$ and $\rho(r)$.
Thus, it forms $T^2(s_i)$ and $T^2(r_i)$ for ${\small s_1,s_2, r_1,r_2,\cdots}$.
To guarantee the effect stutter equivalence,  the sequences
${\small (T^2(s), a_1, T^2(s_1)}$, ${\small a_2, \cdots)}$, denoted by $T^3(s)$, and ${\small  (T^2(r), a_1, T^2(r_1), a_2, \cdots)}$, denoted by $T^3(r)$, should be stutter equivalent.
The process proceeds until ${\small T^k(s) = T^k(r)}$ for any $k$.
Therefore, the trace sets of each state on the paths from $s$ and $r$ will influence the effect equivalence of the starting states $s$ and $r$.
We continue the process, until for any $k$, ${\small T^k(s) = T^k(r)}$.

\subsection{$K$-traces}

From the above discussions, in order to arrive at an adequate notion
of state equivalence to reflect the execution effect, we need to consider not only the traces from $s$
and $r$, but also the traces from each intermediate state that lies on the paths from them. 
This motivates the effect equivalence to be defined in an inductive way, which coincides with the max-trace equivalence \cite{Glabbeek96}.
For ${\small k \in {\mathcal N}}$ and state $s$, let $T^k(s)$ denote the $k$-trace set of $s$.

\begin{definition}[\cite{Glabbeek96}]\label{k-trace}
The notions of $k$-traces and $k$-trace sets of a system $\Delta$ are defined as follows:
\begin{enumerate}
  \item $T^k(s)$ is the set of all $k'$-traces of $s$, for  ${\small k' < k}$.

  \item A $k$-trace of a state $s_0$ is obtained from a sequence ${\scriptsize (T^k(s_0), a_1}$, ${\scriptsize T^k(s_1), a_2\cdots}$, ${\scriptsize a_n}$, ${\scriptsize T^k(s_n))}$
   such that $\Delta$ has a path ${\scriptsize s_0 \xlongrightarrow{a_1} s_1 \xlongrightarrow{a_2}}$ ${\scriptsize \cdots
  \xlongrightarrow{a_n} }$ ${\scriptsize s_n}$,  by replacing all subsequences ${\scriptsize (T^k(s_i)}$, ${\scriptsize a_{i+1},}$ ${\scriptsize T^k(s_{i+1}), a_{i+2},}$ $\cdots,$ ${\scriptsize a_{i+l},}$ ${\scriptsize T^k(s_{i+l}))}$ with ${\scriptsize a_{i+1} =}$ ${\scriptsize a_{i+2} = \cdots = }$ ${\scriptsize a_{i+l}}$ ${\scriptsize = \tau}$, and ${\scriptsize T^k(s_i)}$ ${\scriptsize = }$ ${\scriptsize  T^k(s_{i+1}) = \cdots = }$  ${\scriptsize T^k(s_{i+l})}$ with ${\scriptsize T^k(s_i)}$.

\end{enumerate}
Two states $r$ and $s$ are \emph{$k$-trace equivalent}, written ${\small r \equiv_{k} s}$, if
${\small T^k(r) = T^k(s)}$; They are \emph{max-trace equivalent}, written ${\small r \equiv s}$,
if ${\small r \equiv_{k} s}$ for all $k$.
\qed
\end{definition}

It is straightforward to see that $\equiv_k$ and $\equiv$ are equivalence relations.
By definition, ${\small T^0(s) = \emptyset}$ for every state $s$, and
$T^1(s)$ is just the set of the ordinary traces from $s$; $T^2(s)$ includes all the 0-trace and 1-trace of $s$,
and so on for ${\small T^3(s), \cdots, T^k(s)}$,
which keeps track of more trace information of intermediate states during the execution from $s$.
Also if ${\small k' < k}$ then ${\small T^{k'}(s) \subseteq T^k(s)}$. Note that ${\small r \equiv_{k} s}$ implies ${\small r \equiv_{k'} s}$ for any ${\small k' < k}$.
From this it follows that, for any object system, there exists a
$k$ such that ${\small r \equiv_{k} s}$ iff ${\small r \equiv_{k+1} s}$.
The smallest such a $k$ is called the {\em cap} of the system.

\begin{Expl} \rm \label{example:1}
In Figure~\ref{HW-traces}, $T^2(s)$ and $T^2(r)$ are computed as follows.
$$
{\footnotesize
\begin{array}{lll}
T^2(s)&= & \{(T^1(s),\tau,T^1(s_1),\tau,T^1(s_2),\mathtt{t_1.ret},T^1(s_3),\cdots), \\
&& (T^1(s),\tau,T^1(r),\cdots),\cdots\} \\
T^2(r)&= &\{(T^1(r),\tau,T^1(r_1),\tau,T^1(r_2),\mathtt{t_1.ret},T^1(r_3),\cdots), \\
&&(T^1(r),\tau,T^1(r_7),\tau,T^1(r_8),\mathtt{t_1.ret},T^1(r_9),\cdots),\cdots\}
\end{array}
}
$$
Since ${\scriptsize T^1(s_3) \neq T^1(r_3)}$ ${\scriptsize \neq T^1(r_9)}$, it follows ${\scriptsize T^2(s) \neq T^2(r)}$.
\end{Expl}


\subsection{Effect equivalence and stutter steps}

We define the effect equivalence relation based on $\equiv$.

\begin{definition} \label{effect}
Let ${\small \Delta}$ be an object system.
States $s$ and $r$ in $\Delta$ are effect equivalent if and only if ${\small s \equiv r}$.
\qed
\end{definition}

\begin{definition}
Let ${\small \Delta = (S, \xlongrightarrow{}, {\mathcal A}, s_0)}$ be an object system. There are some notations.
\begin{itemize}
  \item An \emph{effect step} is a path ${\small s_1 \xlongrightarrow{\tau}\cdots }$ ${\small \xlongrightarrow{\tau} s_n \xlongrightarrow{a} r}$ with ${\small s_1 \equiv \cdots \equiv}$ ${\small s_n \not\equiv r}$~${\small(n \geq 1, a \in {\mathcal A})}$, denoted by ${\small\textbf{\textsc{ES}}(s_1,a,r)}$. Sometimes ${\small s \xlongrightarrow{\tau} r}$ with ${\small s \not\equiv r}$ is denoted by ${\small s \xlongrightarrow{\tau}_{\not\equiv} r}$.
  \item An \emph{effect state} is an effect equivalence class ${\small [s]_{\equiv}}$ of $s$, which is defined by ${\small [s]_{\equiv} = \{s' \mid s \equiv s', s' \in S\}}$.
  \item Let ${\small\textbf{\textsc{E}}(\sigma)}$ be the set of effect states on path ${\small\sigma}$.
  \qed
\end{itemize}
\end{definition}

As we mentioned, the effect equivalence relation should satisfy (i) the trace equivalence and (ii) the stutter equivalence w.r.t. the effect.
We define the effect stutter equivalence on paths and prove the result.

\begin{definition} \label{stutter-equivalent}
Let $\sigma$ and $\rho$ be paths.
\begin{enumerate}
  \item They are $k$-trace stutter equivalent, if the $k$-trace obtained from $\sigma$
        is the same as the $k$-trace obtained from $\rho$.
  \item They are effect stutter equivalent, denoted by ${\small \sigma \simeq \rho}$, if they are $k$-trace equivalent for any ${\small k>0}$.
  \qed
\end{enumerate}

\end{definition}

\begin{theorem}\label{thm:11}
For a step ${\small s \xlongrightarrow{\tau} r}$, $s{\small\equiv r}$ iff for any path $\sigma(s)$,
there exists path $\sigma(r)$, such that ${\small\sigma(s)\simeq\sigma(r)}$.
\end{theorem}


By Theorem~\ref{thm:11}, 
the stutter step in object systems can be precisely captured by the relation $\equiv$.

\begin{definition}\label{stutter-steps}
Let ${\small \Delta}$ be an object system.
A internal transition ${\small s \xlongrightarrow{\tau} r}$ is a stutter step in the system ${\small \Delta}$, if and only if ${\small s \equiv r}$.
\qed
\end{definition}

\begin{lemma}\label{lemma-2}
Let $\rho$ be a $\tau$-loop. Any steps on $\rho$ are stutter steps.
\end{lemma}

Let's look back on HW queue again.  In Figure~\ref{HW-traces}, the non-stutter steps are colored red.
(Note that when adding more operations,  more effect steps, e.g., ${\small r_1 \xlongrightarrow{} r_2}$ are exposed.)
We can see ${\small s_3 \xlongrightarrow{} r_3}$ labeled with $\mathtt{t_3.E_2}$ and ${\small s_3 \xlongrightarrow{} s_4}$ labeled with $\mathtt{t_2.D_4}$
are effect steps. Executions ${\small \rho: 0 \xLongrightarrow{} s_9}$ and ${\small \sigma: 0 \xLongrightarrow{} r_6}$ have the same trace to reach $s_4$ and $r_3$ respectively, and $\tau$-paths after $s_4$ and $r_3$ have different effect states. So we can conclude that different return actions on $\rho$ and $\sigma$ is caused by the executions of $\mathtt{t_3.E_2}$ and $\mathtt{t_2.D_4}$ from $s_3$.

\section{The Relation of Effect Race}



Like the concept of the data race, the effect race is also the basic concept in concurrent programs, on which almost all the program analysis and verification
implicitly depend. 
Based on the effect equivalence relation, we define the effect race relation $\ll$ on internal transitions. %

\subsection{A preliminary result}

We first show a preliminary result about the relation of effect steps and executions.
Let $\sigma(s,s')$ denote a path from $s$ to $s'$ on $\sigma$ if ${\small s \neq s'}$; or a single $s$, if ${\small s = s'}$.
The theorem shows that: for any ${\small s \not\equiv_{k} r}$, there are ${\small\sigma(s, s')}$ and ${\small\rho(r, r')}$ that pass along different effect states, but the same trace to reach $s'$ and $r'$ such that ${\small s' \not\equiv_{1} r'}$.

\begin{theorem}\label{thm:6}
Let ${\small \Delta=(S, \rightarrow, {\mathcal A}, s_0)}$ be an object system. For any states ${\small s, r \in S}$, 
if ${\small s \not\equiv r}$, then there exist paths ${\small\sigma(s, s')}$ from $s$ to $s'$ and ${\small\sigma(r, r')}$ from $r$ to $r'$ satisfying:
\begin{enumerate}
  \item ${\small\sigma(s, s')}$ and ${\small\sigma(r, r')}$ have the same trace;
  \item ${\small\textbf{\textsc{E}}(\sigma(s, s')) \cap  \textbf{\textsc{E}}(\sigma(r, r')) = \emptyset}$;
  \item $s' \not\equiv_1 r'$.
\end{enumerate}
\end{theorem}

\begin{Proof}
Let $k$ be the cap of $\Delta$. There exists $l$ with ${\small 1 \leq l \leq k}$ such that ${\small s \not\equiv_l r}$, but ${\small s \equiv_{l-1} r}$.
\emph{Base.} If ${\small l=1}$, let ${\small s_n = s}$ and ${\small r_m = r}$, these results are straightforward.
\emph{Induction.} For ${\small l \geq 2}$, suppose that for any states $s'$ and $r'$ with ${\small s' \not\equiv_{l-1} r'}$, the results 1-3 hold.
Because ${\small s \not\equiv_l r}$, there exits a ${\small(l-1)}$-trace $\rho(s)$ from $s$ such that ${\small \rho(s) \not\in}$  ${\small T^{l}(r)}$,
or there exits a ${\small(l-1)}$-trace $\rho(r)$ from $r$ such that ${\small \rho(r) }$ ${\small \not\in T^{l}(s)}$.
Suppose ${\small \rho(s) \not\in T^{l}(r)}$.
Let ${\small(l-1)}$-trace ${\small \rho(s) =}$ ${\small (T^{l-1}(s), a_1, T^{l-1}(s_1), \cdots)}$ with ${\small a_1 \in {\mathcal A}}$ and ${\small s \not\equiv_{l-1}}$ ${\small s_1}$. Because ${\small s \not\equiv_l r}$ with ${\small l \geq 2}$, we have ${\small s \equiv_1 r}$.
Therefore, there must exist ${\small(l-1)}$-trace ${\small \rho(r) = (T^{l-1}(r),}$ ${\small b_1, T^{l-1}(r_1),}$ ${\small \cdots)}$ such that ${\small a_1 = b_1}$.
Because ${\small \rho(s) \not\in T^{l}(r)}$, it follows ${\small T^{l-1}(s_1)}$ ${\small \neq  T^{l-1}(r_1)}$, that is, ${\small s_1 \not\equiv_{l-1} r_1}$.
By hypothesis, it is easy to see that the results 1-3 hold. The case ${\small \rho(r) \not\in T^{l}(s)}$ can be proved similarly.
\qed
\end{Proof}

Theorem~\ref{thm:6} is a general result regarding the relation of ${\small s \not\equiv r}$ and executions.
In fact, not all of execution fragments satisfying
Theorem~\ref{thm:6} have a meaningful connection with the effect of ${\small s \xlongrightarrow{} r}$.
In Figure~\ref{fig:1} (1) and (2), suppose ${\small s \xlongrightarrow{}_{\not\equiv} r}$, ${\small s \xlongrightarrow{}_{\not\equiv} s'}$ and ${\small s \xlongrightarrow{}_{\not\equiv} s_1 \xlongrightarrow{}_{\not\equiv} s_2}$. It is easy to see $\rho$ from $s$ (blue lines) and $\sigma$ from $r$ (red lines) satisfying Theorem~\ref{thm:6}. But the branch consisting of $\sigma$ and $\rho$ do not identify ${\small s\not\equiv r}$.
In diagram (1), since there are no intermediate states along the paths ${\small s \xlongrightarrow{} s'}$ and ${\small r \xlongrightarrow{} s'}$, by Definition~\ref{k-trace}, ${\small s \equiv r}$. In diagram (2), there are two pathes ${\small s \xlongrightarrow{} s_1 \xlongrightarrow{} s_2}$ and ${\small r \xlongrightarrow{} s_2}$, where $s_1$ is an intermediate state. But there is no branch from $s_1$ to make ${\small s \not\equiv s_1 \not\equiv s_2}$. So, ${\small s \equiv r}$ w.r.t. $\sigma$ and $\rho$.


\begin{figure}[hpbt]
  \centering
  \includegraphics[width=24em]{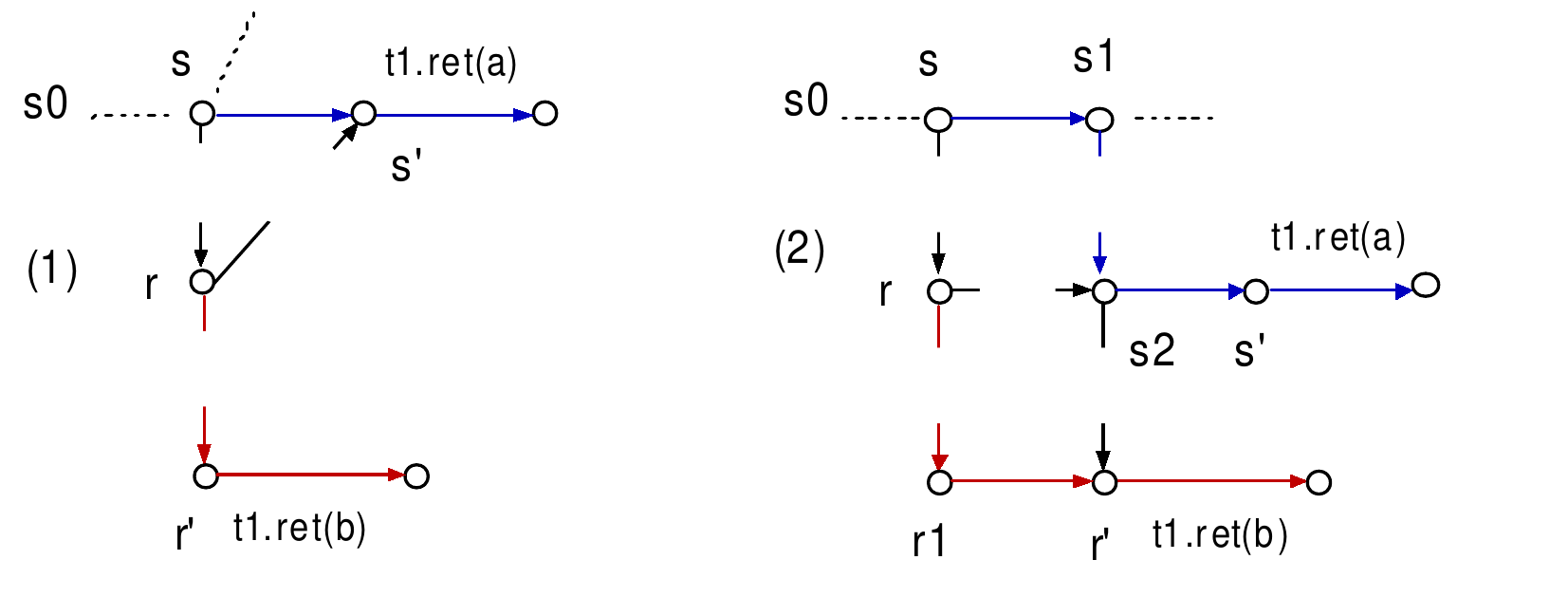}\\
  \caption{Executions and effect steps.}\label{fig:1}
\end{figure}

Hence, it is not proper to understand effect steps from the entire visible actions. We need a precise relation
to reveal the meaning of effect steps for the real programs.

\subsection{Effect race}


\begin{definition}
Let $a$ be a action in object systems. Two effect steps ${\small\textbf{\textsc{ES}}(s,\tau,r)}$ and ${\small\textbf{\textsc{ES}}(s,a,s')}$ are a branch unit, if ${\small s' \not\equiv r}$.
\qed
\end{definition}

\begin{definition}\label{idenitify}
An internal effect step ${\small\textbf{\textsc{ES}}(s,\tau,r)}$ is identified by effect step ${\small\textbf{\textsc{ES}}(s,a,q)}$, where ${\small a}$ is any action in object systems, if
\begin{enumerate}
  \item ${\small\textbf{\textsc{ES}}(s,\tau,r)}$ and ${\small\textbf{\textsc{ES}}(s,a,q)}$ are a branch unit; and
  \item for any ${\small\textbf{\textsc{ES}}(r,a,l)}$, ${\small l\not\equiv q}$.
  \qed
\end{enumerate}
\end{definition}

In Figure~\ref{fig:1} (1),  ${\small s \xlongrightarrow{} r}$ is not identified by ${\small s \xlongrightarrow{} s'}$.

\begin{lemma}\label{lemma-1}
For each internal effect step $\alpha$, there exists an effect step $\beta$, such that
$\alpha$ is identified by $\beta$.
\end{lemma}

Let ${\small s \xLongrightarrow{\tau} r}$ denote zero or more $\tau$-steps, and ${\small s \xLongrightarrow{a} r}$ denote  ${\small s \xLongrightarrow{\tau} \xlongrightarrow{a} \xLongrightarrow{\tau} r}$.

\begin{definition} \label{independence}
Let ${\small \Delta = (S, \xlongrightarrow{}, {\mathcal A}, s_0)}$ be an object system, and ${\small\textbf{\textsc{ES}}(s,\tau,r)}$ be identified by ${\small\textbf{\textsc{ES}}(s,a,q)}$.
\begin{enumerate}
  \item ${\small\textbf{\textsc{ES}}(s,\tau,r)}$ is independent of ${\small\textbf{\textsc{ES}}(s,a,q)}$ w.r.t. effect state $[l]_{\equiv}$, if
        (i) there exist ${\small r \xLongrightarrow{a} l_1}$ and ${\small q \xLongrightarrow{\tau} l_2}$ such that ${\small l_1,l_2 \in [l]_{\equiv}}$ for any ${\small(a \in {\mathcal A})}$ or (ii) there is ${\small q \xLongrightarrow{\tau} l_2}$ such that ${\small l_2 \equiv r}$ for ${\small a=\tau}$.
  \item ${\small\textbf{\textsc{ES}}(s,\tau,r)}$ is dependent of ${\small\textbf{\textsc{ES}}(s,a,q)}$ w.r.t.  $[l]_{\equiv}$, if it is not independent of ${\small\textbf{\textsc{ES}}(s,a,q)}$ w.r.t. $[l]_{\equiv}$.
  \qed
\end{enumerate}
\end{definition}

\begin{definition}[The effect race relation] \label{effectraceI}
Let $\Phi$ be a set of effect steps and $\ll \subseteq \Phi \times \Phi$ be the effect race relation.
For effect steps $\alpha, \beta \in \Phi$, ${\small \alpha \ll \beta}$ iff $\alpha$ is dependent of $\beta$ w.r.t. any effect states.
\qed
\end{definition}


\begin{theorem}
Let $\alpha$ and $\beta$ be internal effect steps.
The following properties of $\ll$ hold:
\begin{enumerate}
  \item[(1)] symmetric: ${\small\alpha \ll  \beta}$ implies  ${\small \beta \ll \alpha}$.
  \item[(2)] irreflexive:${\small \alpha \not\ll \alpha}$.
  \item[(3)] non-transitive: ${\small \alpha \ll \beta}$ and ${\small \beta \ll \mu}$ does not imply ${\small \alpha \ll \mu}$.
\end{enumerate}
\end{theorem}

\begin{Expl}
In Figure~\ref{HW-traces}, ${\small s \xlongrightarrow{} r}$ is independent of ${\small s \xlongrightarrow{}}$ ${\small s_2}$ w.r.t. $[r_2]_{\equiv}$,
and ${\small s_3 \xlongrightarrow{} r_3 \ll s_3 \xlongrightarrow{} s_4}$.
\qed
\end{Expl}

As we mentioned in Section 1, for the object program,  call and return actions do not access the shared object state.
So the effect race is only related to internal steps.

\begin{lemma} \label{lemma-4}
Let ${\small \Delta = (S, \xlongrightarrow{}, {\mathcal A}, s_0)}$ be an object system, and ${\small\textbf{\textsc{ES}}(s,\tau,r)}$ is identified by ${\small\textbf{\textsc{ES}}(s,a,q)}$.
\begin{enumerate}
  \item If $a$ is a visible action, then ${\small\textbf{\textsc{ES}}(s,\tau,r)}$ is independent of ${\small\textbf{\textsc{ES}}(s,a,q)}$ w.r.t. some effect state.
  \item If ${\small\textbf{\textsc{ES}}(s,\tau,r)\ll\textbf{\textsc{ES}}(s,a,q)}$, then ${\small a = \tau}$.
\end{enumerate}
\end{lemma}


\begin{theorem}\label{thm:5}
For each effect step $\alpha$, there must exist an effect step $\beta$
such that
\begin{enumerate}
  \item either $\alpha$ is independent of $\beta$;
  \item or ${\small \alpha \ll \beta}$.
\end{enumerate}
\end{theorem}

The relation $\ll$ is defined on general internal executions, where an effect step is a path consisting of several stutter steps $l_1 \xLongrightarrow{\tau} l_n$ and a step ${\small l_n \xlongrightarrow{}_{\not\equiv} l'}$. By the stutter equivalence in Definition~\ref{stutter-equivalent}, in fact,
each effect step $\alpha$ is stutter equivalent to an internal transition ${\small s \xlongrightarrow{\tau} r}$. Therefore, for ${\small \alpha \ll \beta}$, there exist
non-stutter steps ${\small s \xlongrightarrow{\tau} r}$ and ${\small s \xlongrightarrow{\tau} l}$ such that ${\small s \xlongrightarrow{\tau} r \ll s \xlongrightarrow{\tau} l}$. 

\begin{lemma} \label{lemma-7}
For ${\small \alpha \ll \beta}$ in $\Delta$, there exist ${\small s \xlongrightarrow{\tau}_{\not\equiv} r}$ and ${\small s \xlongrightarrow{\tau}_{\not\equiv} l}$ such that ${\small s \xlongrightarrow{\tau}_{\not\equiv} r \ll s \xlongrightarrow{\tau}_{\not\equiv} l}$ in $\Delta$.
\end{lemma}

\begin{definition} \label{effectraceII}
Let $\alpha$ and $\beta$ be labeled with instructions $c_1$ and $c_2$ respectively.
If ${\small \alpha \ll \beta}$, then $c_1$ and $c_2$ are effect race instructions from $s$, denoted by
${\small c_1 \ll_s c_2}$.
\qed
\end{definition}


\subsection{Effect structures and critical steps}

A neat structure, called \emph{effect structure}, is defined.
We show that each effect step has at least one effect structure that is responsible for recognizing its effect.
For convenience sake, we give the following notations.

\begin{enumerate}
  \item  ${\small\textbf{\textsc{ES}}(s,\tau,r)^{\sigma}}$ denotes that execution $\sigma$ passes through ${\small\textbf{\textsc{ES}}(s,\tau,r)}$, and the two notions
  ${\small\textbf{\textsc{ES}}(s,\tau,r)^{\sigma}}$ and ${\small\textbf{\textsc{ES}}(s,\tau,q)^{\rho}}$ mean that $\sigma$ and $\rho$ share the same prefix from initial $s_0$ to $s$.
  \item  $\sigma' = \sigma(s'/s)$ is an execution, which has the same states of $\sigma$ except replacing the state $s$ in $\sigma$ by $s'$.
\end{enumerate}

\begin{definition}[race structures]\label{effect-race-str}
Let $\Delta$ be an object system.
If ${\footnotesize\textbf{\textsc{ES}}(s_i,\tau,r_i) \ll \textbf{\textsc{ES}}(s_i,\tau,s_{i+1})}$,
the set of executions $\sigma$ and $\rho$ with ${\small\textbf{\textsc{ES}}(s_i,\tau,r_i)^{\sigma}}$ and ${\small\textbf{\textsc{ES}}(s_i,\tau,s_{i+1})^{\rho}}$ in $\Delta$ is called the race structure of the relation $\ll$, denoted by ${\small \textsc{RaceStr}(s_i,r_i,s_{i+1})}$.
\qed
\end{definition}

\begin{figure}[htpb]
  \centering
  \includegraphics[width=24em]{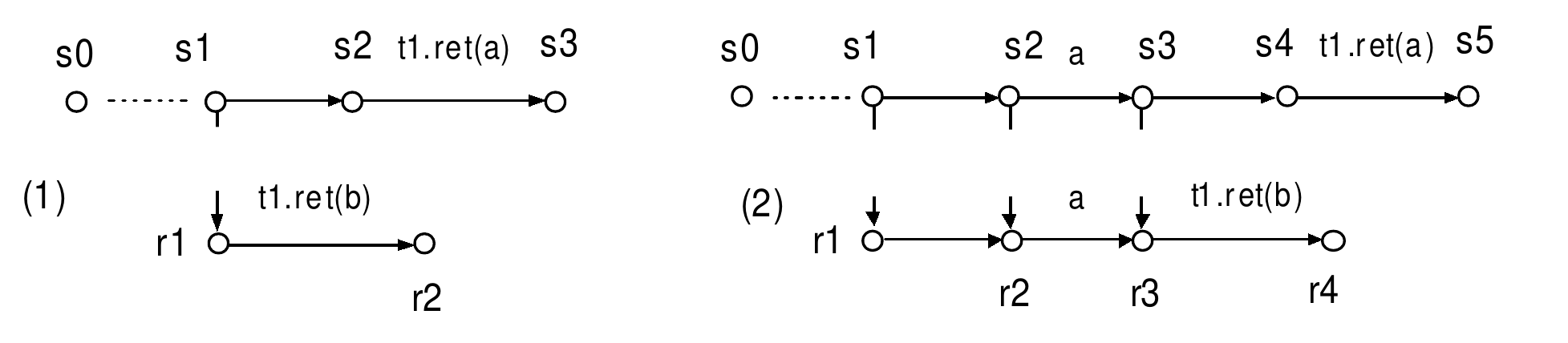}\\
  \caption{Effect structures.}\label{fig:4}
\end{figure}

In Figure~\ref{fig:4} (1), we have ${\scriptsize s_1 \xlongrightarrow{} r_1 \ll s_1 \xlongrightarrow{} s_2}$, so the set of $\sigma$ and $\rho$ is the race structure ${\small \textsc{RaceStr}(s_1,r_1,s_2)}$. The race structure is also called the effect structure, denoted by ${\small \textsc{Estr}(s_1,r_1)}$.  By the symmetry of $\ll$, it is easy to see ${\small \textsc{Estr}(s_1,r_1)}$ ${\small = \textsc{Estr}(s_1,s_2)}$.

Further, in Figure~\ref{fig:4} (2),  we have ${\scriptsize s_3 \xlongrightarrow{} r_3 \ll s_3 \xlongrightarrow{} s_4}$.
Let ${\scriptsize \sigma,\rho \in}$ ${\small \textsc{RaceStr}(s_3,r_3,s_4)}$, where ${\scriptsize \sigma= s_0 \xLongrightarrow{} s_1 \xlongrightarrow{} s_2 }$ ${\scriptsize \xlongrightarrow{} s_3}$ ${\scriptsize \xlongrightarrow{} r_3 \xlongrightarrow{} r_4}$ and ${\scriptsize \rho = s_0 \xLongrightarrow{} s_1  \xlongrightarrow{} s_2}$ ${\scriptsize \xlongrightarrow{} s_3 \xlongrightarrow{} s_4}$ ${\scriptsize \xlongrightarrow{} s_5}$. Let ${\scriptsize \sigma' =\sigma(r_2/s_3)}$, that is, ${\scriptsize \sigma' = s_0}$ ${\scriptsize \xLongrightarrow{} s_1 \xlongrightarrow{} s_2}$ ${\scriptsize \xlongrightarrow{} r_2}$ ${\scriptsize \xlongrightarrow{} r_3 \xlongrightarrow{} r_4}$.
If ${\scriptsize s_2 \xlongrightarrow{} r_2}$ is independent of ${\scriptsize s_2 \xlongrightarrow{} s_3}$ w.r.t. ${\scriptsize [r_3]_{\equiv}}$, then the set ${\small \{\sigma'\} \cup \textsc{Estr}(s_3, r_3)}$ is the effect structure ${\small \textsc{Estr}(s_2, r_2)}$.
So ${\small \textsc{Estr}(s_1, r_1)}$ can be defined. Obviously ${\small \textsc{Estr}(s_3, r_3)}$ ${\small \subseteq\textsc{Estr}(s_2, r_2)}$ ${\small \subseteq \textsc{Estr}(s_1, r_1)}$.

\begin{definition}[effect structures] \label{estr}
Let $\Delta$ be an object system.
An effect structure of ${\footnotesize\textbf{\textsc{ES}}(s_i,\tau,r_i)}$, denoted by ${\small\textsc{Estr}(s_i,r_i)}$, is a set of executions, which is defined as:
\begin{enumerate}
  \item If there exists ${\footnotesize\textbf{\textsc{ES}}(s_i,\tau,r_i) \ll \textbf{\textsc{ES}}(s_i,\tau,s_{i+1})}$, then ${\small \textsc{Estr}(s_i,r_i)}$ ${\small = \textsc{RaceStr}(s_i,r_i,s_{i+1})}$.
  \\ \vspace{-0.7em}
  \item Let ${\small\rho,\sigma \in \textsc{Estr}(s_{i},r_{i})}$ with ${\small\textbf{\textsc{ES}}(s_{i},\tau,r_{i})^{\sigma}}$ and \\ ${\small\textbf{\textsc{ES}}(s_{i}, a_{i},}$ ${\small s_{i+1})^{\rho}}$ and ${\small\textbf{\textsc{ES}}(s_{i-1},a_{i-1},s_{i})^{\rho}}$, if \\
  ${\footnotesize \textbf{\textsc{ES}}(s_{i-1},\tau,r_{i-1})} {\mbox{ is independent of }} {\footnotesize \textbf{\textsc{ES}}(s_{i-1},a_i,s_{i})}
  \\  {\small \mbox{ w.r.t. } } {\small [r_{i}]_{\equiv}}$, \\
  then ${\small \textsc{Estr}(s_{i-1},r_{i-1})}$ = ${\small \{\sigma(r_{i-1}/s_{i})\}\cup \textsc{Estr}(s_{i},r_{i})}$. 
\end{enumerate}
\qed
\end{definition}

An important property of ${\small\textsc{Estr}(s,r)}$ is that, ${\small s \not\equiv r}$ decided in the entire system $\Delta$ can be precisely decided in ${\small\textsc{Estr}(s,r)}$. If the set of executions in $\Delta$ is a subset of the set of executions in $\Delta'$, then we denote ${\small \Delta \subseteq \Delta'}$.
The "precisely" means that: for any larger systems ${\small\Delta' \subseteq \Delta}$, which has the same effect states as ${\small\textsc{Estr}(s,r)}$, but with more transitions, there still has ${\small s \not\equiv r}$ in $\Delta'$. This shows that ${\small s \not\equiv r}$ always holds, which do not need other branch and extra intermediate effect states outside ${\small\textsc{Estr}(s,r)}$ to decide. Note that effect states and related transitions in ${\small\textsc{Estr}(s,r)}$ and $\Delta'$ come from $\Delta$. 

\begin{theorem} \label{thm:2}
States ${\small s \not\equiv r}$ in ${\small\textsc{Estr}(s,r)}$.
\end{theorem}

\begin{theorem} \label{thm:7}
Let $\Delta$ be an object system. 
For $\Delta'$ with ${\small \textsc{Estr}(s,r)}$ ${\small \subseteq \Delta'}$ ${\small \subseteq \Delta}$, if $\Delta'$ has the same effect states as ${\small\textsc{Estr}(s,r)}$, then ${\small s \not\equiv r}$ in $\Delta'$.
\end{theorem}

An effect step may have more than one effect structures depending on how many branch units associated with the step.
We show the existence of the effect structure for each effect step.
The result can be proved by Theorem~\ref{thm:5} and Definition~\ref{estr}.

\begin{lemma}\label{thm:8}
Let $\Delta$ be a finite object system. For each ${\small s_1 \xlongrightarrow{\tau}_{\not\equiv} r_1}$ in $\Delta$, there exists ${\small\textsc{Estr}(s_1,r_1)}$.
\end{lemma}

Now the precise connection between effect steps and executions is established by means of effect structure ${\small \textsc{Estr}(s,r)}$.
The following theorem improves the results of Theorem~\ref{thm:6} by restricting executions to ${\small \textsc{Estr}(s,r)}$.


\begin{theorem}\label{thm:9}
Let $\Delta$ be a finite system. For each step ${\small s \xlongrightarrow{\tau}_{\not\equiv} r}$,
there are executions $\sigma$ and $\rho$ and states $s'$ and $r'$ such that
\begin{enumerate}
  \item ${\small \sigma, \rho \in \textsc{Estr}(s,r)}$;
  \item ${\small\rho(s, s')}$ and ${\small\sigma(r, r')}$ have the same race;
  \item ${\small\textbf{\textsc{E}}(\rho(s, s')) \cap  \textbf{\textsc{E}}(\sigma(r, r')) = \emptyset}$;
  \item $s' \not\equiv_1 r'$.
\end{enumerate}
\end{theorem}

By Theorem~\ref{thm:9},  
it is easy to see that each ${\small s \xlongrightarrow{}_{\not\equiv} r}$ in ${\small \textsc{Estr}(s,r)}$ has the potential to represent a race with another non-stutter step in ${\small \textsc{Estr}(s,r)}$ to cause different traces on $\sigma$ and $\rho$. Such the race can be seen immediately when ${\small \textsc{Estr}(s,r)}$ is a race structure.

\begin{definition}[critical steps] \label{cs}
A $\tau$-transition ${\small s \xlongrightarrow{\tau} r}$ is called a critical step of an object system, if ${\small s \not\equiv r}$.
\qed
\end{definition}

Let $\textsc{R}$ be a race structure of $\Delta$, and ${\small \textsc{RaceStr}(\Delta) = \bigcup \textsc{R}}$ denote all the race structures in $\Delta$.

\begin{theorem}\label{thm:4}
Let ${\footnotesize \Delta}$ be a finite object system and
${\small \Phi_{\Delta}}$ be the set of critical steps in $\Delta$.
$$
{\small
\begin{array}{l}
\Phi_{\Delta} =  {\scriptsize\bigcup\limits_i}\{s \xlongrightarrow{\tau} r \mid \textsc{Estr}(s,r) \supseteq \textsc{R}_i, \textsc{R}_i \in \textsc{RaceStr}(\Delta) \}
\end{array}
}
$$
\end{theorem}

Therefore, all critical steps in a system can be found based on the race structure. On the other hand, finding the race structure can fall back on each critical step.


\section{Three Effect Theorems}


In the following, we consider any completed concurrent histories with the same method calls but different return actions.
These concurrent histories are the main concerns for the verification.

\begin{definition}
Let $H_1$ and $H_2$ be completed concurrent histories with the same call actions.
If they have different return actions, then ${\scriptsize H_1 \neq H_2}$; otherwise, ${\scriptsize H_1 = H_2}$.
\end{definition}

In an object system $\Delta$, there may have many interleaved executions such that their visible actions are the same as $H_1$ or $H_2$.
We focus on all the executions $\sigma$ in $\Delta$ such that $H(\sigma) = H_1$ or $H(\sigma) = H_2$.
These interleaved executions constitute a subsystem of $\Delta$ relevant with the visible actions of $H_1$ and $H_2$, denoted by ${\scriptsize \Delta{(H_1H_2)}}$.

\begin{definition}
Let $\Delta$ be an object system and ${\scriptsize H_1 \neq H_2}$. The subsystem
${\scriptsize \Delta(H_1H_2)} = \{\sigma \mid \sigma \mbox{ is an execution of } \Delta$ \\
$\mbox{ s.t. } {\scriptsize H(\sigma) =H_1} \mbox{ or } {\scriptsize H(\sigma) = H_2 } \}$.
\qed
\end{definition}

\begin{lemma}\label{lemma-5}
Let $\Delta$ be a finite system. For each ${\small \sigma \in}$ ${\small \textsc{Estr}(s,r)}$,
there is  ${\small \sigma' \in \textsc{RaceStr}(\Delta)}$ s.t. ${\scriptsize H(\sigma) = H(\sigma')}$.
\end{lemma}
\begin{Proof}
By Definitions~\ref{estr} and \ref{independence} and Theorem~\ref{thm:9}.
\end{Proof}

In the following, we give three effect theorems about the race structures and histories.

\underline{We first show that:} for two completed concurrent histories ${\small H_1 \neq H_2}$,  $H_1$ and $H_2$ are enumerable by means of the effect race relation in system $\Delta(H_1H_2)$.

\paragraph*{\textsf{\small \underline{Effect Theorem I}:}}

\begin{theorem} \label{thm:12}
If ${\scriptsize H_1 \neq H_2}$, then there exist a race structure $\textsc{R}$ in ${\scriptsize \Delta(H_1H_2)}$
and $\sigma, \rho \in \textsc{R}$ such that ${\scriptsize H(\sigma) = H_1}$ and ${\scriptsize H(\rho) = H_2}$.
\end{theorem}

\begin{Proof}
Since ${\small H_1 \neq H_2}$, there are different return actions ${\small\mathtt{t,ret(a)}}$ on $H_1$ and ${\small \mathtt{t,ret(b)}}$ on $H_2$ for the same method call by $\mathtt{t}$. Therefore, there is a $\tau$-step ${\small s \xlongrightarrow{\tau} r}$ such that ${\small s \not\equiv r}$ is recognized by ${\small\mathtt{t,ret(a)}}$ and
${\small\mathtt{t,ret(b)}}$. By Theorem~\ref{thm:8}, there is a ${\small \textsc{Estr}(s,r)}$ in ${\small \Delta(H_1H_2)}$.
By Theorem~\ref{thm:9} and Lemma~\ref{lemma-5}, there are ${\small \sigma, \rho \in \textsc{R}}$ such that ${\small H(\sigma) = H_1}$ and ${\small H(\rho) = H_2}$.
\qed
\end{Proof}

In ${\scriptsize \Delta(H_1H_2)}$, there may have more than one race structures according to the event orders on histories.
Each race structure is associated with two internal steps $\alpha$ and $\beta$ such that they satisfy ${\small \alpha \ll \beta}$.
Therefore, different return actions of $H_1$ and $H_2$ are in essence caused by the race on precedence orders of the steps  $\alpha$ and $\beta$.

To validate the application of Theorem~\ref{thm:12} in infinite systems,
\underline{we show that:} the effect race relation in a small system also holds in lager systems with more method calls.
Therefore, the effect race relation in finite systems are the sound basis to analyze algorithm and prove the correctness (e.g., lineraizability) of infinite systems.

\begin{lemma} \label{lemma-8}
Let $\Delta'$ be an object system, and ${\small \Delta(H_1H_2)}$ ${\small \subseteq \Delta'}$. If ${\small s \xlongrightarrow{\tau}_{\not\equiv} r}$ in $\Delta(H_1H_2)$, then ${\small s \xlongrightarrow{\tau}_{\not\equiv} r}$ in $\Delta'$.
\end{lemma}
\begin{Proof}
Because ${\small s \xlongrightarrow{\tau}_{\not\equiv} r}$ in $\Delta(H_1H_2)$,
by Definition~\ref{k-trace}, there is an effect step ${\scriptsize s \xLongrightarrow{}}$ ${\scriptsize\xlongrightarrow{\tau}_{\not\equiv} s_1}$ on an execution $\sigma$ with ${\scriptsize H(\sigma) = H_1}$ (or $H_2$), such that for any path ${\scriptsize r \xLongrightarrow{} r_1}$ in ${\scriptsize  \Delta(H_1H_2)}$, ${\small s_1 \not\equiv r_1}$.
Suppose ${\scriptsize s \equiv r}$ in $\Delta'$. Then there exists ${\scriptsize r \xLongrightarrow{} l}$ on ${\scriptsize \rho' \not\in \Delta(H_1H_2)}$ such that ${\scriptsize l \equiv s_1}$ in ${\scriptsize \Delta'}$. By the stutter equivalence in Theorem~\ref{thm:11}, ${\scriptsize H(\rho') = H(\sigma)}$. Therefore, ${\scriptsize \rho'\in\Delta(H_1H_2)}$, which is a contradiction.
\qed
\end{Proof}

\paragraph*{\textsf{\small \underline{Effect Theorem II}:}}

\begin{theorem} \label{thm:10}
Let ${\scriptsize \Delta(H_1H_2) \subseteq \Delta'}$.
$$
\begin{array}{l}
{\small c_1 \ll_s c_2} \mbox{ in } {\scriptsize \Delta(H_1H_2)} \mbox{ implies } {\small c_1 \ll_s c_2} \mbox{ in } {\Delta'}
\end{array}
$$
\end{theorem}
\begin{Proof}
By Lemma~\ref{lemma-8}.
\end{Proof}

For a system $\Delta$, let ${\small H(\Delta) = \{H \mid \exists H'. H \neq H' \mbox{ in } \Delta\}}$ be a set of different concurrent histories in $\Delta$.
Although a fine-grained program involves a large number of disordered concurrent histories, \underline{we show that:} for the entire object program $\Delta$,
completed concurrent histories in ${\small H(\Delta)}$ are enumerable in race structure ${\small \textsc{RaceStr}(\Delta)}$.

\paragraph*{\textsf{\small \underline{Effect Theorem III}:}}

\begin{theorem}\label{thm:3}
Let $\Delta$ be an object system. 
$$
{
\begin{array}{l}
 {H}(\Delta) \subseteq \{ H(\sigma) \mid \sigma \in \textsc{RaceStr}(\Delta)\}
\end{array}
}
$$
\end{theorem}
\begin{Proof}
By Theorem~\ref{thm:12} and Theorem~\ref{thm:10}.
\qed
\end{Proof}

Therefore, the effect race relation is the accurate relation to capture the internal steps, of
which precedence orders cause disorder concurrent histories.
Verifying histories in the entire system can thus be transformed to verify the simple race structure ${\small\textsc{RaceStr}(\Delta)}$.

\section{Branching Bisimulation}

Branching bisimulation~\cite{Glabbeek96} refines Milner's weak bisimulation~\cite{DBLP:books/daglib/0067019}
by requiring two related states that should preserve not only their
own branching structure but also the branching potentials of all
intermediate states that are passed through. It has been shown that
branching bisimulation is an equivalent characterization of the max-trace equivalence.
Thus, we can provide an efficient way to compute the effect race relation of algorithms.

\subsection{Branching bisimulation for concurrent objects}

\begin{definition}\label{co-sim}
Let ${\scriptsize \Delta = (S, \xlongrightarrow{}, {\mathcal A}, s_0)}$ be an object system. A symmetric relation ${\mathcal R}$ on $S$ is a branching
bisimulation
if for all $(s_1, s_2) \in {\mathcal R}$, the following holds:
\begin{enumerate}
  \item if ${\footnotesize s_1 \xlongrightarrow{a} s_1'}$ where
  $a$ is a visible action,
  then there exists $s_2'$ such that
  ${\footnotesize s_2 \xlongrightarrow{a} s_2'}$ and ${\footnotesize (s_1', s_2')\in \mathcal{R}}$.
  \\[-.5em]
  \item[{\it 2.}] if ${\footnotesize s_1 \xlongrightarrow{\tau} s_1'}$, then either ${(s_1', s_2)\in {\mathcal R}}$,
  or there exist $l_1, \cdots, l_i$, $i \geq 0$, and $s'_2$ such that
  $
  {\footnotesize s_2 \xlongrightarrow{\tau} l_1 \xlongrightarrow{\tau} \cdots }$ ${\footnotesize \xlongrightarrow{\tau} l_i}{\footnotesize \xlongrightarrow{\tau} s_2'}
  $
  and
  ${\footnotesize (s_1, l_1) \in {\mathcal R}}, \cdots, {\footnotesize (s_1, l_i) \in {\mathcal R}}$, \\
  ${\footnotesize (s_1', s_2')\in {\mathcal R}}$.
\end{enumerate}

Let $\approx \DEF \bigcup \{ {\mathcal R} \mid {\mathcal R} \mbox{ is a branching bisimulation}\}$ be the largest branching bisimulation.
Then $\approx$ is an equivalence relation.
\qed
\end{definition}


\begin{theorem}\cite{Glabbeek96} \label{thm:1}
For any states $s$ and $r$ in an object system, $s \equiv r$ if and only if $s \approx r$.
\end{theorem}


For finite state systems, branching bisimulation can be computed in
polynomial time  \cite{Groote90,DBLP:conf/tacas/GrooteW16}.

\subsection{Quotient Object Systems}

Given an object system ${\footnotesize\Delta =(S, \xlongrightarrow{}, {\mathcal A}, s_0)}$, for any $s \in S$, let $[s]_{\approx}$ be the
equivalence class of $s$ under $\approx$, and
$S/\!\!\approx = \{[s]_{\approx}\!\! \mid \!\!s \!\in \!S\}$  the
set of the equivalence classes under $\approx$.

\begin{definition}[Quotient transition system] \label{quotient}
For an object system ${\footnotesize\Delta =(S, \xlongrightarrow{}, {\mathcal A}, s_0)}$,
the quotient transition system $\Delta/ {\approx}$ is defined as:
${\scriptsize
   \Delta / {\approx} = (S/ {\approx}, \xlongrightarrow{}_{\approx}, Act,}$  ${\scriptsize [s_0]_{\approx})
}$,
where the transition relation
$\xlongrightarrow{}_{\approx}$ is generated by the following rules:
  $$
  {\footnotesize
  \begin{array}{ll}
  (1) \frac{\displaystyle s \xlongrightarrow{a} s' }
  {\displaystyle [s]_{\approx} \xlongrightarrow{a}_{\approx} [s']_{\approx}} \ (a \neq \tau) ~~~~
  (2) \frac{\displaystyle s \xlongrightarrow{\tau} s' }
  {\displaystyle [s]_{\approx} \xlongrightarrow{\tau}_{\approx} [s']_{\approx}} \ ((s, s') \not\in \approx)
  \end{array}
  }
  $$
  \qed
\end{definition}

\begin{theorem} \label{thm-quo}
For a $\tau$ path ${\small s_1 \xlongrightarrow{\tau} \cdots \xlongrightarrow{\tau} s_{n-1} \xlongrightarrow{\tau} s_n}$ ${\small\xlongrightarrow{\tau} r}$ in $\Delta$, it is an effect step  if and only if $[s_n]_{\approx} \xlongrightarrow{\tau}_{\approx} [r]_{\approx}$ is a transition in $\Delta/ {\approx}$, where ${\small s_1, \cdots, s_{n-1} \in [s_n]_{\approx}}$.
\end{theorem}


\section{Effects of Concrete Algorithms}

This section takes the real object algorithms as examples to show that complicated executions and their effects, which are informally described and used in the existed work,
can be precisely captured by the effect equivalence relation and the effect race relation.

\subsection{CCAS}

CCAS in Figure~\ref{fig:ccas_orig} is a simplified RDCSS \cite{Victor08} and contains complicated executions.
Instead of returning true or false in conventional $\mathtt{cas}$, the $\mathtt{cas}$
in CCAS returns the old value of the shared variable $\mathtt{a}$. To update the value of $\mathtt{a}$, the thread
first constructs a descriptor with its id $\mathtt{cid}$ and the expected old
value $\mathtt{o}$ and the new value $\mathtt{n}$. The element in $\mathtt{a}$
can be either a value or a descriptor. If the thread read a descriptor
by the cas operation, this means another thread has registered itself
first and a help method called Complete is performed to help that
thread to finish updating the value. Furthermore, a global variable
$\mathtt{flag}$ can also influence the success of Complete.
Initially, ${\small \mathtt{flag} := \mathtt{true}}$ and ${\small\mathtt{a} :=1}$.
We consider an object system $\Delta$ involving the following four concurrent method calls:
${\scriptsize \mathtt{t_1.CCAS(1,2)}}$, $\mathtt{t_2.CCAS(2,3)}$, $\mathtt{t_3.CCAS(2,5)}$ and $\mathtt{t_4.SetFlag(false)}$.
Since an object system enumerates all the possible interleaved steps of each thread at each state, the state space is exponential increase.

\begin{figure}[ht!]
\begin{narrow}{-1ex}{-2ex}
\begin{minipage}{.24\textwidth}
\begin{lstlisting}[basicstyle=\scriptsize,escapechar=|]
C1 CCAS(o, n) {	
C2  local r, d;		
C3  d := cons(cid, o, n);	
C4  |\setlength{\fboxsep}{1pt}\lcolorbox{yellow}{r := cas(\&a, o, d);}|		
C5  while(IsDesc(r)) {
C6   Complete(r);	
C7   |\setlength{\fboxsep}{1pt}\lcolorbox{yellow}{r := cas(\&a, o, d);}|
C8  }		
C9  if(r = o) Complete(d);	
C10 return r; }				
\end{lstlisting}
\end{minipage}
\begin{minipage}{.24\textwidth}
\begin{lstlisting}[basicstyle=\scriptsize,escapechar=|]
C11 Complete(d) {
C12  local b;
C13  |\setlength{\fboxsep}{1pt}\lcolorbox{yellow}{b := flag;}|
C14  if (b)
C15   |\setlength{\fboxsep}{1pt}\lcolorbox{yellow}{cas(\&a, d, d.n);}|
C16  else
C17   |\setlength{\fboxsep}{1pt}\lcolorbox{yellow}{cas(\&a, d, d.o);}|
C18 }

F1 SetFlag(b){|\setlength{\fboxsep}{1pt}\lcolorbox{yellow}{flag := b;}|}
\end{lstlisting}
\end{minipage}
\end{narrow}
\vspace{-2ex}
\caption{The algorithm of CCAS.}\label{fig:ccas_orig}\vspace{-1ex}
\end{figure}

\begin{table}[htbp]
\renewcommand\arraystretch{1.5}
{\scriptsize
\centering
\begin{tabular}{|c|c|c|l|}
\hline
\hline
CCAS  & \#states & \#$\tau$  & Instructions of critical steps \\
\hline
System $\Delta$  & 4382 & 8218    &  $\mathtt{C_4}$ and $\mathtt{C_7}$: when ${\small\mathtt{r} == \mathtt{true}}$; \\
\cline{1-3}
Quo. $\Delta/\!\!\approx$ & 330 & 220  & $\mathtt{C_{15}}$ and $\mathtt{C_{17}}$: when ${\small\mathtt{a} == \mathtt{d}}$; $\mathtt{C_{13}}; \mathtt{F_1}$\\
\hline
\end{tabular}
\caption{Instructions of critical steps for CCAS.}\label{CCAS}
}
\end{table}

The generated system is shown in Table~\ref{CCAS}, where the state spaces of ${\scriptsize \Delta}$ and ${\scriptsize\Delta/\!\!\approx}$ are 4382 and 330 respectively;
the total number of $\tau$-transitions in ${\scriptsize \Delta}$ and ${\scriptsize \Delta/\!\!\approx}$ are 8218 and 220 respectively.

The corresponding instructions of critical steps in ${\scriptsize \Delta}$ are also recoded. Each critical step is an essential state transformation annotated in the proof~\cite{Victor08,Feng13}.

\subsubsection*{Effect equivalence of CCAS}


We first apply the effect equivalence relation to analyzing CCAS.
The main feature of CCAS is the helping, of which the effect is informally described in many verification work (e.g.,\cite{feng12}).
We show effect equivalence relation $\equiv$ precisely captures the implicit meaning of helping.

\begin{figure}[htpb]
  \centering
  \includegraphics[width=24em]{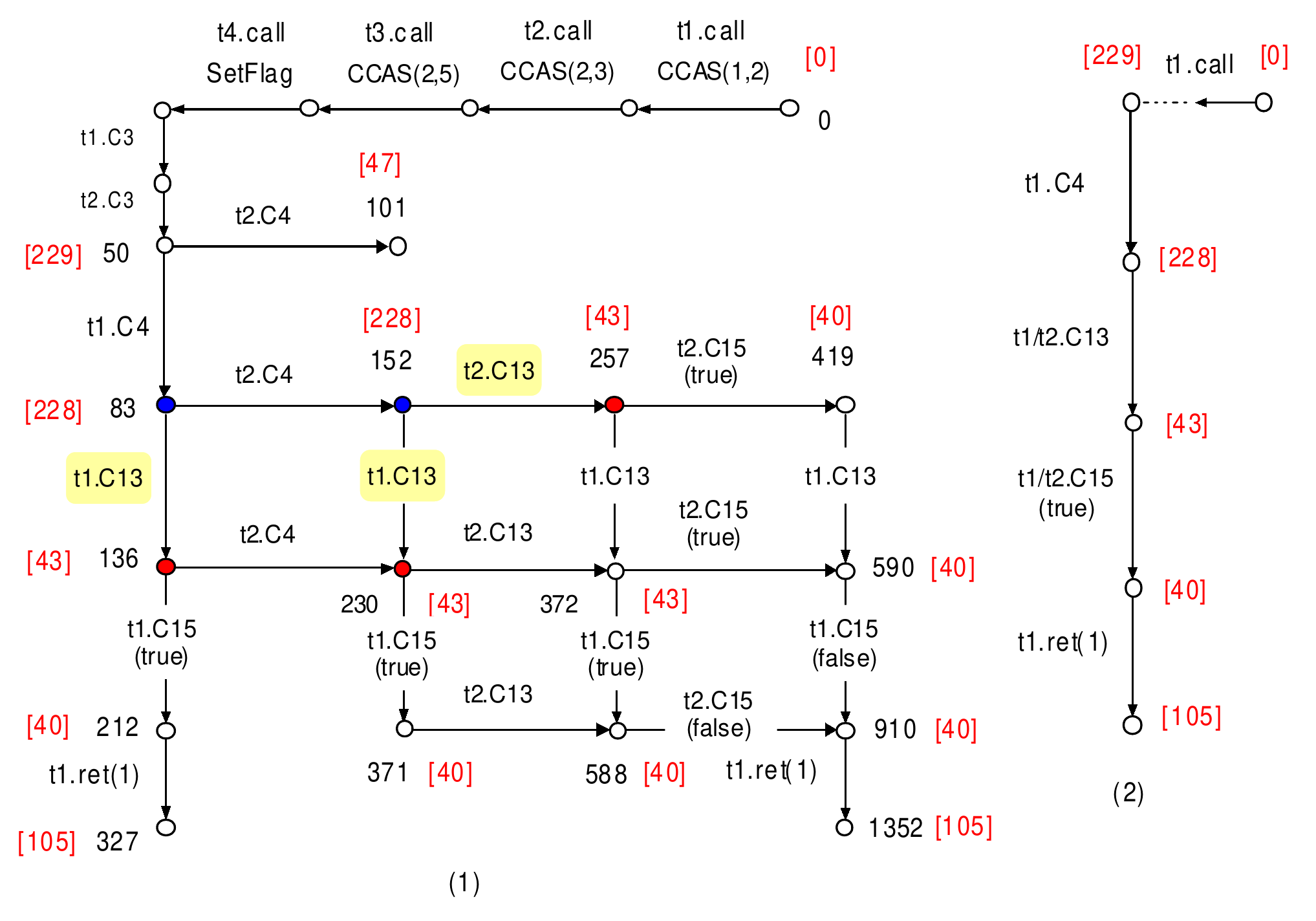}\\
  \caption{(1) Effect equivalence relation of CCAS; (2) Transitions in the quotient.}\label{CCAS-exec}
\end{figure}

The effect equivalence class of states in the entire system $\Delta$ has been computed. Figure~\ref{CCAS-exec} (1) presents
executions of threads ${\small\mathtt{t_1}}$ and ${\small\mathtt{t_2}}$, and marks the state number and equivalence class numbers of each state\footnote{State numbers and
equivalence class numbers are generated by CADP.}, e.g.,
the equivalence class of state 50 is $[229]$. From the equivalence class, it is clear which steps are stutter, and which steps are critical.
In particular, the states of the following $\tau$-steps:
$$
{\small
\begin{array}{l}
 83 \xlongrightarrow{\mathtt{t_1.C_{13}}} 136, \qquad  152 \xlongrightarrow{\mathtt{t_1.C_{13}}} 230, \qquad 152 \xlongrightarrow{\mathtt{t_2.C_{13}}} 257
\end{array}
}
$$
correspond to the same effect states ${\small [228]}$ and ${\small [43]}$, implying ${\small \mathtt{t_2.C_{13}}}$ and ${\small \mathtt{t_1.C_{13}}}$ on these transitions take the same effect.
The analysis of executions in Figure~\ref{CCAS-exec} (1) can be equivalently transformed to analyzing the quotient in Figure~\ref{CCAS-exec} (2), where $\mathtt{t_1.C_{13}}$ and $\mathtt{t_2.C_{13}}$ (and $\mathtt{C_{15}}$) share the same transition.
This clearly shows thread $\mathtt{t_2}$ helps thread $\mathtt{t_1}$ complete the method call before completing its own method call.


\subsubsection*{Effect race of CCAS}

We now see the effect race relations of CCAS. The quotient is partly shown in Figure~\ref{CCAS-qo}, where the corresponding instructions of critical steps are labeled at each step.
There are the following effect race relation:
$$
{\small
\begin{array}{l}
\mathtt{t_1.C_{13}} \ll_{187} \mathtt{t_4.F_1} \qquad \mathtt{t_1.C_{15}} \ll_{91} \mathtt{t_2.C_{17}}
\end{array}
}
$$

One race is about shared variable $\mathtt{flag}$, where different orders of reading $\mathtt{flag}$ by $\mathtt{t_1.C_{13}}$ and updating ${\small\mathtt{flag}}$ by $\mathtt{t_4.F_1}$  will result in different return actions.
The other race is about changing variable $\mathtt{a}$, where the effect race of $\mathtt{t_1.C_{15}}$ and $\mathtt{t_2.C_{17}}$ appear at the state $[91]$ where $\mathtt{flag}$ has been assigned to $\mathtt{false}$. If $\mathtt{t_2.C_{17}}$ takes effect, then $\mathtt{t_2}$ helps $\mathtt{t_1}$ complete the method call and keep the old value of $\mathtt{a}$ unchanged. Otherwise, if $\mathtt{t_1.C_{15}}$ takes effect first,  then $\mathtt{t_1}$ updates $\mathtt{a}$ to the new value $2$ since $\mathtt{t_1}$ reads $\mathtt{flag:=true}$ earlier. 

The quotient lets us quickly find the effect race relation.
This example confirms \textsf{\small \underline{Effect Theorem I}} that effect race instructions are accurate to cause all different visible actions.
By \textsf{\small \underline{Effect Theorem II}}, these effect race relations in Figure~\ref{CCAS-qo} is valid for analyzing larger systems.

\begin{figure}[htpb]
  \centering
  \includegraphics[width=22em]{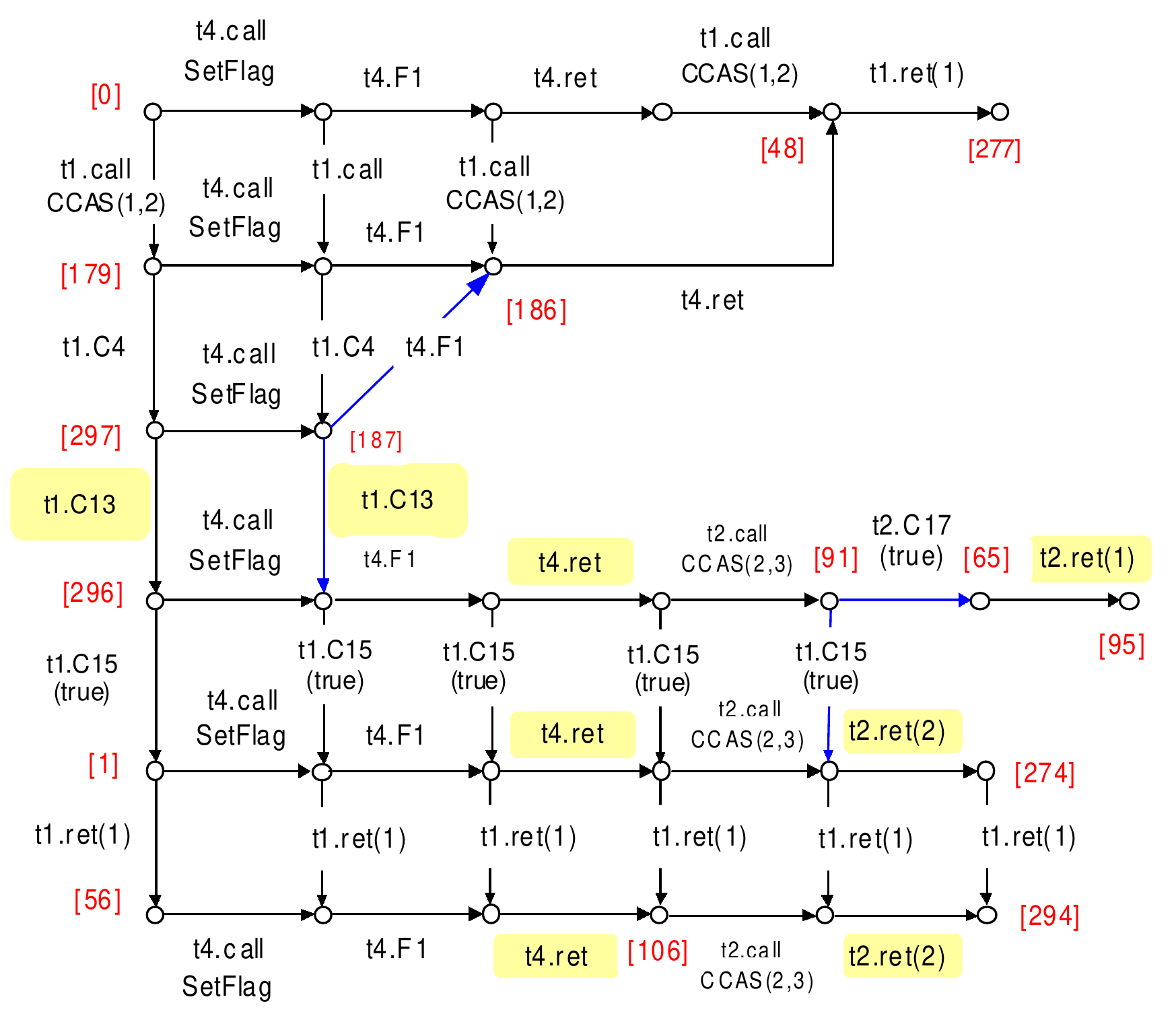}\\
  \caption{Effect structures and effect races in $\Delta/\!\!\approx$.}\label{CCAS-qo}
\end{figure}

\vspace{-1em}

\subsection{The MS lock-free queue}

A fine-grained program may involve many instructions to modify the shared state.
We take MS queue~\cite{DBLP:conf/podc/MichaelS96} to show that not all the instructions access to the shared variable have the potential to generate the effect race relation.
Figure~\ref{pic-1} shows the implementation of the methods $\mathtt{enq}$ and $\mathtt{deq}$ of the MS queue.
The queue's representation a linked-list, where $\mathtt{Head}$ and $\mathtt{Tail}$ refer to the first and the last node respectively.

\begin{figure}[ht]
 \begin{minipage}{.25\textwidth}
 \vspace{-2ex}
 \begin{lstlisting}[basicstyle=\scriptsize,escapechar=|]
 01Enq(v) {
 02 local x,t,s,b;
 03 x:=new_node(v);
 04 while(true) {
 05  t:=Tail; s:=t.next;
 06  if (t=Tail) {
 07   if (s=null) {
 08   |\setlength{\fboxsep}{1pt}\lcolorbox{yellow}{b:=cas(\&(t.next),s,x);}|
 09    if (b) {
 10     cas(&Tail,t,x);
 11     return true; }
 12   }else cas(&Tail,t,s);
 13  }
 14 }
 15} \end{lstlisting}
 \end{minipage}
 \begin{minipage}{.22\textwidth}
 \begin{lstlisting}[basicstyle=\scriptsize,escapechar=|]
 16Deq() {
 17 local h,t,s,v,b;
 18 while(true) {
 19 h:=Head; t:=Tail;
 20 |\setlength{\fboxsep}{1pt}\lcolorbox{yellow}{s:=h.next;}|
 21 |\setlength{\fboxsep}{1pt}\lcolorbox{yellow}{if (h==Head);}|
 22 if (h=t) {
 23  if(s=null)
 24      return EMPTY;
 25    cas(&Tail,t,s);
 26   }else {
 27     v :=s.val;
 28     |\setlength{\fboxsep}{1pt}\lcolorbox{yellow}{b:=cas(\&Head,h,s);}|
 29     if(b) return v;}
 30 }
 31}\end{lstlisting}
 \end{minipage}
 \vspace{-2ex}
 \caption{MS lock-free queue: enqueue and dequeue} \label{pic-1} \vspace{-.5ex}
 \end{figure}

Let $\Delta$ be an object system including 2 threads invoking methods for 3 times.
The state spaces of $\Delta$ and $\Delta/\!\!\approx$, and instructions of critical steps are shown in Table~\ref{MS}, from which we can see that all the steps labeled with ${\small\mathtt{Tail}}$ are stutter steps. Therefore, instead of manually analyzing intricate executions, we can compute which instructions access to the shared state are critical.
\vspace{-.5em}
\begin{table}[htbp]
\renewcommand\arraystretch{1.5}
{\scriptsize
\centering
\begin{tabular}{|c|c|c|l|}
\hline
\hline
MS queue  & \#states & \#$\tau$  & Instructions of critical steps \\
\hline
$\Delta$  & 49038 & 72950   &  $\mathtt{Line_8}$ and $\mathtt{Line_{28}}$: when ${\small\mathtt{cas} = \mathtt{true}}$;  \\
\cline{1-3}
$\Delta/\!\!\approx$ & 863 & 448  & $\mathtt{Line_{21}}$: when ${\small\mathtt{h} == \mathtt{Head}}$; $\mathtt{Line_{20}}$\\
\hline
\end{tabular}
\caption{Instructions of critical steps for MS queue.}\label{MS}
}
\end{table}

\vspace{-2em}

\section{Linearization Points}

An informal description of LPs \cite{Herlihy08} is shown as follows:
every method call on an execution appears to take effect instantly at some time point between its invocation and its response, behaving
as defined by the sequential definition. Such the point, corresponding to the execution of an instruction, is referred to as the LP of the method call.
In the following, we give a precise definition of LPs. 

Let $\mathtt{e_i}$ be an operation associated with thread $\mathtt{t_i}$, and $\mathtt{e_i.m}$ denote the invoked method $\mathtt{m}$ in $\mathtt{e}$.
For convenience, each operation in an execution has a different name.
Let ${\small \mathtt{S}}$ be a sequential history including operations
${\small \mathtt{e_0 <_{S} e_1 <_{S} \cdots <_{S} e_n}}$. The partial order of
the invocation and response events and $\tau$-step $\alpha$ on an execution $\sigma$ can be denoted as ${\small \mathtt{e.call <_{\sigma} \alpha <_{\sigma} }}$ ${\small \mathtt{e.ret <_{\sigma} \cdots}}$.

\begin{definition} \label{LP}
Let $\Delta$ be an object system, and ${\small \mathtt{S_1}}$ and ${\small \mathtt{S_2}}$ legal sequential histories including operations
${\small \mathtt{e_0 <_{\mathtt{S_1}} \cdots e_{i-1} <_{\mathtt{S_1}} e_i}}$ and
${\small \mathtt{e_0 <_{\mathtt{S_2}} \cdots e_{i-1} <_{\mathtt{S_2}} e_{i+1}}}$.
If there exist executions $\sigma$ and $\rho$ in $\Delta$, and steps $\alpha$ labeled with $\mathtt{t_i.c}$ and $\beta$ with $\mathtt{t_{i+1}.c'}$ such that
\begin{enumerate}
  \item ${\small H(\sigma) \sqsubseteq_{\textsf{lin}} \mathtt{S_1}}$ and ${\small H(\rho) \sqsubseteq_{\textsf{lin}} \mathtt{S_2}}$;
  \item ${\small \alpha \ll \beta}$;
  \item ${\scriptsize\mathtt{e_k.ret} <_{\sigma} \alpha}$ and ${\scriptsize \mathtt{e_k.ret} <_{\rho} \beta}$ ~ ${\footnotesize (0 \leq k \leq i-1)}$.
\end{enumerate}
then, instruction $\mathtt{c}$ is an LP for $\mathtt{e_i.m}$ on $H(\sigma)$.
\qed
\end{definition}

By the symmetry of $\ll$, the instruction labeled on $\beta$ is an LP for $\mathtt{e_{i+1}.m}$ on $H(\rho)$.
It is easy to see that the effect race of $\alpha$ and $\beta$ represents the race on completing operations ${\mathtt{e_i}}$ and ${\mathtt{e_{i+1}}}$ on $\sigma$ and $\rho$, behaving as defined in the specification. Let ${\small c_t}$ be the number of critical steps associated with thread $t$.

\begin{definition}
Let $\Delta$ be an object system and $\mathtt{m}$ be an object method of $\Delta$.
\begin{enumerate}
  \item The LP of ${\small\mathtt{m}}$ is non-fixed, if there is a $\small{\mathtt{(t,call,m)}}$ in $\Delta$, such that between the $\small{\mathtt{(t,call,m)}}$ and the matched ${\small \mathtt{(t,ret,m)}}$, ${\small c_t > 1}$.
  \item The LP of method ${\small\mathtt{m}}$ is fixed, if it is not non-fixed for any object system $\Delta'$ with ${\small \Delta \subseteq \Delta'}$.
  \qed
\end{enumerate}
\end{definition}

In practice, to see the LP of a method call, a larger system is needed to reveal all the effect race relations of the method call.
For example, in Figure~\ref{HW-traces}, when threads invoke 2 times of methods, the effect race relation ${\small \mathtt{t_3.E_2} \ll_s \mathtt{t_1.E_2}}$ from $s$ is exposed. It is easy to see $\mathtt{E_2}$ is the LP for ${\small\mathtt{Enq(b)}}$ on the execution ${\small s_0 \xLongrightarrow{} s \xlongrightarrow{} r \xLongrightarrow{} r_6}$. Other algorithms can also be analyzed by Definition~\ref{LP}.

\section{A Quantitative Analysis of Fine-Grained Algorithms}

Understanding the fine-grained algorithm is
difficult due to a lot ofintricate interleavings. Instead of manual analysis, the critical steps of an algorithm can be computed.
The critical steps-rate, shorted as C-rate, is given as follows:

\begin{quote}
  C-rate = $\displaystyle \frac{\mbox{the number of $\tau$-transitions in $\Delta/\!\!\approx$}} {\mbox{the number of $\tau$-transitions in $\Delta$}}$
\end{quote}
\vspace{.5em}
In general, for different algorithms with the same parameters of method calls,
more critical steps an algorithm has, more complicated interleavings the algorithm involves.
In the section, we give a quantitative analysis for different fine-grained implementation in terms of critical steps.
All experiments are conducted on a server which is equipped with a ${\small 4\times 12}$-core AMD CPU @ 2.1
GHz and 192 GB memory under 64-bit Debian 7.6.

\subsection{Critical steps of Herlihy and Wing queue}

The branching bisimulation quotient of the HW queue in Figure~\ref{HW-traces}
is shown in Table~\ref{HW-quotient}, where the object system in Figure~\ref{HW-traces}
has $292$ states and $670$ transitions (among them $368$ $\tau$-transitions);
the quotient system has only $52$ states and $116$ transitions (among them $28$ $\tau$-transitions).
The C-rate is $7.6\%$ ($28$ out of $368$), which implies only a small portion
of the invisible steps in the original system are responsible for the effect of executions, while the remaining
$92.4\%$ are stutter hence can be abstracted away, as is done in the quotient system.


\begin{table}[htbp]
\renewcommand\arraystretch{1.2}
{\scriptsize
\centering
\begin{tabular}{|c|c|c|c|}
\hline
\hline
HW queue        & \#states    & \#total trans. & \# $\tau$-trans. \\
\hline
System $\Delta$ & 292 & 670 & 368 \\
\hline
Quo. $\Delta/\!\!\approx$       & 52  & 116 & 28 \\
\hline
C-rate &-&- &7.6\%  \\
\hline
\end{tabular}
\caption{{\small The C-rate of the HW queue in Figure~\ref{HW-traces}.}}\label{HW-quotient}
}
\end{table}


Table~\ref{k-trace-HW} summaries the $\not\equiv$-transitions in the quotient. There are 24 $\tau$-transitions ${\small [s]_{\approx} \xlongrightarrow{\tau}_{\approx} [r]_{\approx}}$ that $s$ and $r$ are not 1-trace equivalent, which correspond to the instructions
$E_1$ (4), $E_2$ (16) and $D_4$ (4); and 4 $\tau$-transitions ${\small [s]_{\approx} \xlongrightarrow{\tau}_{\approx} [r]_{\approx}}$ that $s$ and $r$ are not 2-trace equivalent but 1-trace equivalent, which are labeled with $E_2$. Any transition labeled with $D_2$ are stutter steps.

\begin{table}[htbp]
\renewcommand\arraystretch{1.2}
{\scriptsize
\centering
\begin{tabular}{|c|c|c|c|c|c|}
\hline
\hline
$k$-trace inequiv.    & \#$\tau$-trans. in $\Delta/\!\!\approx$    & $E_1$ & $E_2$ & $D_2$ & $D_4$\\
\hline
$\not\equiv_1$      & 24    & 4 & 16 & 0 & 4 \\
\hline
$\equiv_1$ but $\not\equiv_2$  & 4   & 0 & 4 & 0 & 0  \\
\hline
\end{tabular}
\caption{{\small HW queue: the instructions of critical steps.}}\label{k-trace-HW}
}
\end{table}

Therefore, we can see that almost all the instructions that access to the shared state in the Herlihy and Wing queue are critical, which cause the complicated races.
The red lines in Figure 2 are critical steps saved in $\Delta/\!\!\approx$.

\subsection{Critical steps of various algorithms}

We compute the C-rates of various algorithms, and show that the finite system with 2 or 3 threads are enough to reveal all the algorithm essentials.
Table~\ref{tab:rate} shows the number of $\tau$-transitions of object system $\Delta$ and $\Delta/\!\!\approx$ and computes the C-rate.  Table~\ref{tab:turning points} summaries the corresponding instructions of critical steps.
All the quotients are computed in a few seconds, and all the instructions of critical steps (c.f. Table~\ref{tab:turning points}) are the essential instructions that are used in existed theorem proofs (e.g.,\cite{Feng13,spin09}).
These experimental results allow us to analyze and compare different fine-grained implementation in a quantitative way.
In the following, we combine Tables~\ref{tab:rate} and \ref{tab:turning points} together to analyze the queue and list.

The MS queue and DGLM queue contain two methods $\mathtt{Enq}$ and $\mathtt{Deq}$. Although their implementation are different, with the same scale of method calls, they have the same quotient (i.e., 448 in Table~\ref{tab:rate}), and the same instructions of critical steps shown in Table~\ref{tab:turning points}.  All instructions are related to the access to either $\mathtt{t.next, h.next}$, or ${\mathtt{Head}}$. The same quotient and instructions of the two queues gives a hint that proof techniques for MS and DGLM should be the same.
The Herlihy and Wing queue has been analyzed in Section 9.1. 

The HM (Harris-Michael) list is lock-free, and the lazy list and optimistic list are implemented based on fine-grained locks. The synchronization primitive, e.g., $\mathtt{lock}$, as critical steps is preserved in the quotient.
Under the same parameters of threads and method calls, the lazy list has the largest number of critical steps (15297), and optimistic list has the smallest number of critical steps (9843). From Table~\ref{tab:turning points}, each method of optimistic list contains only one instruction $\mathtt{lock}$, but the other two lists contain more instructions. This indicates that the essential interleavings of the optimistic list are much more simple than other two lists, accordingly, the proof should also be easier for optimistic lists.

\begin{table}[htbp]
\renewcommand\arraystretch{1.2}
{\scriptsize
\centering
\begin{tabular}{|c|c|c|c|c|c|c|}
\hline
{\scriptsize\#Th-Op.} & {\scriptsize Objects} & {\scriptsize \#$\tau$ in $\Delta$}& {\scriptsize \#$\tau$ in $\Delta/\!\!\approx$} & {\scriptsize Time(s)}& {\scriptsize C-rate} \\
\hline
2-3 & MS & 72950  & 448  & 0.23 &0.61\% \\
\hline
2-3 & DGLM  & 62328 & 448  & 0.27 &  0.72\% \\
\hline
3-2& HW &128727& 4062  & 0.48 &  3.2\%  \\
\hline
3-2& HM list &1007592 & 11385 & 1.71  &  1.1\%  \\
\hline
3-2& lazy list & 2607504 & 15297 & 5.92 &  0.59\%    \\
\hline
3-2& opt. list &  2670636  & 9843 &  4.38 & 0.37\%    \\
\hline
4-1& CCAS &2296 & 115  & 0.07 &   5.0\% \\
\hline
2-2& HP &25366 & 108 & 0.11 &  0.43\% \\
\hline
\end{tabular}
\caption{{\small State space and C-rates of systems in Table \ref{tab:turning points}.}}\label{tab:rate}
}
\end{table}

\begin{table*}[htpb]
\centering
\newcommand{\tabincell}[2]{\begin{tabular}{@{}#1@{}}#2\end{tabular}}
\scalebox{1}[1]{
\begin{tabular}{|c|c|l|c|c|c|c|c|c|c|}
\hline
\hline
  \multirow{2}{*}{\scriptsize\#Th-Op.} & \multirow{2}{*}{{\footnotesize Algorithms}} & \hspace{1cm}\multirow{2}{*}{{\footnotesize The corresponding instructions of critical steps for various algorithms}} \\[1.2em]
\hline
  {\scriptsize 2-3} & {\scriptsize MS/DGLM queue~\cite{DBLP:conf/podc/MichaelS96}} &  \tabincell{l}{\scriptsize $\mathtt{\textbf{Enq}}$: $\mathtt{b:=cas(\&(t.next), s, x)}$, when $\mathtt{cas}$ is $\mathtt{true}$.  \\ \scriptsize $\mathtt{\textbf{Deq}}$: (1) $\mathtt{s := h.next}$; ~ (2) $\mathtt{if (h==Head)}$ when it is $\mathtt{true}$;  ~ (3) $\mathtt{b:=cas(\&Head, h, s)}$ when $\mathtt{cas}$ is $\mathtt{true}$. }  \\
\hline
  {\scriptsize 3-2} & {\scriptsize HW queue~\cite{Herlihy90}} &  \tabincell{l}{ \scriptsize $\mathtt{\textbf{Enq}}$: (1) $\mathtt{(i,back):=(back,back+1)}$; ~ (2) $\mathtt{AR[i]:=x}$. \\
  \scriptsize $\mathtt{\textbf{Deq}}$: $\mathtt{(x, AR[i]):=(AR[i], null)}$, when $\mathtt{x := AR[i] ~ (i != null)}$ or
                                                                   \scriptsize  $\mathtt{x := AR[i] ~ (i == null)}$. } \\
\hline
  {\scriptsize 3-2} & {\scriptsize HM list~\cite{Herlihy08}} &  \tabincell{l}{\scriptsize $\mathtt{\textbf{Add}}$:  \scriptsize $\mathtt{pred.next.cas(curr,node,false,false)}$ when $\mathtt{cas}$ is $\mathtt{true}$. ~  \\
  \scriptsize $\mathtt{\textbf{Rem}}$:  \scriptsize $\mathtt{curr.next.cas(succ,succ,false,true)}$ when $\mathtt{cas}$ is $\mathtt{true}$.  \\
  \scriptsize $\mathtt{\textbf{Find}}$: (1) $\mathtt{curr=pred.next.getReference()}$;  ~ (2) $\mathtt{succ=curr.next.get(marked)}$, when $\mathtt{marked==false}$.}  \\
\hline
  {\scriptsize 3-2} & {\scriptsize Lazy list~\cite{DBLP:journals/ppl/HellerHLMSS07}} & \tabincell{l}{\scriptsize $\mathtt{\textbf{Add}}$:  \scriptsize $\mathtt{pred.next=node}$; ~ $\mathtt{lock}$;
  \scriptsize $\mathtt{\textbf{Rem}}$:  \scriptsize $\mathtt{curr.marked=true}$; ~ $\mathtt{lock}$;  \\
  \scriptsize $\mathtt{\textbf{Contains}}$: \scriptsize (1) $\mathtt{curr=curr.next}$; (2) $\mathtt{curr.marked = false}$}\\
\hline
  {\scriptsize 3-2} & {\scriptsize  Opt. list~\cite{Herlihy08}} & \tabincell{l}{\scriptsize $\mathtt{\textbf{Add}}$:  \scriptsize $\mathtt{lock}$.
  \scriptsize $\mathtt{\textbf{Rem}}$:  \scriptsize  $\mathtt{lock}$.
  \scriptsize $\mathtt{\textbf{Contains}}$: \scriptsize  $\mathtt{lock}$.}\\
\hline
  {\scriptsize 4-1} & {\scriptsize CCAS~\cite{DBLP:conf/popl/TuronTABD13}} &  \tabincell{l}{\scriptsize $\mathtt{\textbf{CCAS}}$: (1) $\mathtt{r:=cas(\&a,o,d)}$; ~ (2)
  $\mathtt{b:=flag}$; ~ \scriptsize (3) $\mathtt{cas(\&a,d,d.n)}$ and (4) $\mathtt{cas(\&a,d,d.o)}$ when $\mathtt{cas}$ succeeds. \\ \scriptsize $\mathtt{\textbf{SetFlag}}$: $\mathtt{flag:=b}.$} \\
\hline
  {\scriptsize 2-2} & {\scriptsize {HP(Treiber)}~\cite{Michael04}} &  \tabincell{l}{\scriptsize $\mathtt{\textbf{Pop}}$:
(1) $\mathtt{cas(\&Top,old,x)}$ when $\mathtt{cas}$ succeeds; (2) $\mathtt{old:=Top}$  \quad \scriptsize $\mathtt{\textbf{Push}}$: $\mathtt{cas(\&Top,old,x)}$ when $\mathtt{cas}$ succeeds}  \\
\hline
\end{tabular}}
\caption{{\small The instructions of critical steps in fine-grained algorithms computed by the branching bisimulation quotient.}}\label{tab:turning points}
\vspace{.6ex}
\end{table*}

For the HM list, the critical steps of methods $\mathtt{Add}$ and $\mathtt{Rem}$ are labeled by successful $\mathtt{cas}$, which implies an item are successfully added to or removed from the list. However the critical steps of unsuccessful $\mathtt{Add}$ and $\mathtt{Rem}$ contains two instructions, which are in the while-loops of $\mathtt{Find}$. Which steps labeled with these instruction are critical steps depends on the concrete execution. 
The $\mathtt{Add}$ and $\mathtt{Rem}$ of lazy list have the similar analysis as HM list, except it has more steps labeled with $\mathtt{lock}$, which makes the number of critical steps of lazy list is larger than that of HM list (11385). Method $\mathtt{Contains}$ of the lazy list also has two instructions.  More than one instructions in  a method
implies the method has non-deterministic effects, making verification more difficult.



\begin{table}[htpb]
\renewcommand\arraystretch{1.2}
{\footnotesize
\begin{tabular}{|p{0.05cm}<{\centering}|p{0.1cm}<{\centering}|*{7}{p{0.7cm}<{\centering}|}}
	   	\multicolumn{2}{c}{} & \multicolumn{1}{c}{MS} & \multicolumn{1}{c}{DGLM}& \multicolumn{1}{c}{HW} & \multicolumn{1}{c}{Opt} & \multicolumn{1}{c}{Lazy} & \multicolumn{1}{c}{HML} & \multicolumn{1}{c}{HP}\cr
		\hline		
		\multirow{4}{*}{2} &
			 2 & 1.2\% & 1.3\% & 4.7\%  & 1.06\% & 1.5\% & 3.4\% & 0.43\%\cr\cline{2-9}
			 & 3 & 0.61\% & 0.72\% & 3.8\%  & 1.08\% & 1.4\% & 2.7\% & 0.21\%\cr\cline{2-9}
			 & 4 & 0.32\% & 0.39\% & 3.1\% & 1.08\% & 1.1\% & 1.9\% & 0.07\%\cr\cline{2-9}
			 & 5 & 0.16\% & 0.20\% & 2.0\%  & 1.08\% & 0.94\% & 1.5\% & 0.01\%\cr
		\hline 
		\multirow{1}{*}{3} &
			3 & 0.04\% & 0.06\% & 1.47\% & 0.30\% & 0.36\% &  0.5\% & ${\scriptsize 0.03}$\%$^{*}$\cr
		\hline
\end{tabular}\caption{{\small The C-rate in different concrete algorithms.}}\label{tab:TPrate}\vspace{-1ex}
}
\end{table}
A comparison of different finite instances of these algorithms are summarized in Table~\ref{tab:TPrate}.
For most algorithms, the C-rates are less than 2\%. 
Because the quotient abstracts away all $\tau$-transitions irrelevant to the execution effect, it shows that the enormous state space can be obtained based on quotients.
Furthermore, from Table~\ref{tab:TPrate}, we can see that if there are more threads with more operations, the C-rates will
become less and less for scalable concurrent data structures.




\section{Related Work and Conclusions}


A plethora of proof-based techniques has been developed based on rely-guarantee reasoning~(e.g.,\cite{DBLP:conf/ifip/Jones83,Victor08,Victor10,Feng13,DBLP:conf/esop/KhyzhaDGP17}) or simulation methods~(e.g.,\cite{Schellhorn12,DBLP:journals/tocl/SchellhornDW14,Colvin06,Derrick11}) to verify concurrent objects.
These techniques often involve identifying LPs and their auxiliary variables to construct the state function~\cite{DBLP:journals/tcs/AbadiL91}. However, although these work are applicable to a wide range of popular non-blocking algorithms (e.g., \cite{DBLP:journals/ppl/HellerHLMSS07,Herlihy90,DBLP:conf/popl/DoddsHK15}), they lack a formal basis for understanding fine-grained concurrency. 
Due to the intricate executions, analyzing the fine-grained interleavings puzzles verifiers when conducting a proof.

Our work provides a formal and feasible basis for this issue. Effect equivalence relation and effect race relation are proved accurate to explain various phenomenon in fine-grained concurrency (Section 7).
Effect theorems reveal that effect race relation is the accurate relation to capture the internal instructions, of which different execution orders cause chaotic histories. Since the effect equivalence relation in finite systems can be computed by the branching bisimilar in the polynomial time, these results can be used efficiently in practice.

Model-based verification work of the fine-grained concurrency have also been proposed in e.g., \cite{Liu13,Alur10,spin09,pldi10,arXiv16,arXiv17}.
These work can verify and debug linearizability of finite systems automatically. But how to correct understanding non-blocking algorithms is still obscure
for these verification work, some of which also involves on manually annotated LPs (e.g., \cite{spin09}).
Understanding and debugging finite concurrent systems can facilitate proofs of infinite systems~\cite{cegar,book08}.
Effect theorem II in our paper shows that the effect race relations of a small system still hold in larger system.
This implies that essential effect relations exposed on a small system are the sound base for analyzing infinite system.
This paper does not discuss how to select the smallest finite system to reveal all the essential effect relations for the inductive proof of infinite systems. This will be the future work.

Various weak bisimulation~\cite{DBLP:books/daglib/0067019,Namjoshi97,DBLP:conf/icalp/BaetenG87,Glabbeek96} have been proposed in process algebra. Nothing but branching (or stuttering) bisimulation satisfies the stuttering equivalence that is an important condition in our paper to apply bisimulation to analyzing concurrent programs. \\

\noindent
\emph{\textbf{Conclusions.}} This paper attempts to provide a formal and efficient basis for analyzing fine-grained algorithms. Two basic concepts -- the effect equivalence
relation and effect race relation -- are defined to precisely capture
various phenomena of effects in concurrent programs, which are obscure and intricate for programmers and verifiers.
A lot of interleavings with instructions access to the shared states make understanding fine-grained algorithms difficult. Effect
theorems reveal that chaotic concurrent histories are in essence caused by the internal steps satisfying
the effect race relation.  This validates the accuracy and application of the effect race relation in practice, which provides verifiers a clear clue to analyze complex algorithms. Further, linearization points are characterized by the effect race relation.
We have conducted a lot experiments to show the efficiency of these definitions for analyzing real fine-grained concurrent programs.


\subsection*{Acknowledgement}
The author would like to thank Huimin Lin and Joost-Pieter Katoen for a lot of discussions on the previous work~\cite{arXiv16,arXiv17}. The ideas of using the $k$-trace in \cite{Glabbeek96} and computing the state-space reduction factor in Section 9 are suggested by Huimin Lin.

\bibliographystyle{abbrvnat}

\begin{thebibliography}{00}

\bibitem{DBLP:books/daglib/0067019}
Robin Milner. 1989. {\em Communication and {C}oncurrency}. Prentice Hall.

\bibitem{cadp}
Hubert Garavel,  Fr{\'{e}}d{\'{e}}ric Lang, Radu Mateescu, and Wendelin Serwe. 2013. {CADP} 2011: a toolbox for the construction and analysis of distributed processes. In {\em STTT}, vol.15, 2, 89-107. 

\bibitem{DBLP:journals/tocl/SchellhornDW14}
Gerhard Schellhorn, John Derrick, and Heike Wehrheim. 2014. A Sound and Complete Proof Technique for Linearizability of Concurrent Data Structures. {\em {ACM} Trans. Comput. Log.} 15. 4 (2014), 31:1--31:37.


\bibitem{DBLP:conf/popl/DoddsHK15}
Mike Dodds, Andreas Haas, and Christoph M. Kirsch. 2015. A Scalable, Correct Time-Stamped Stack. In {\em POPL 2015}. 233-246.
 

\bibitem{Michael04}
Maged M. Michael. 2004. Hazard {P}ointers: {S}afe {M}emory {R}eclamation for {L}ock-{F}ree {O}bjects. {\em {IEEE} Trans. Parallel Distrib. Syst.} 15, 6 (2004), 491-504.


\bibitem{DBLP:conf/podc/MichaelS96}
Maged M. Michael and Michael L. Scott. 1996. Simple, {F}ast, and {P}ractical {N}on-{B}locking and {B}locking {C}oncurrent {Q}ueue {A}lgorithms. In {\em PODC 1996}, 267-275.



\bibitem{DBLP:conf/popl/TuronTABD13}
Aaron Joseph Turon, Jacob Thamsborg, Amal Ahmed, Lars Birkedal, and
Derek Dreyer. 2013. Logical {R}elations for {F}ine-{G}rained {C}oncurrency
In {\em POPL 2013}, ACM, 343--356.


\bibitem{Herlihy08}
Maurice Herlihy and Nir Shavit. 2008. {\em The {A}rt of {M}ultiprocessor {P}rogramming.} Morgan Kaufmann.


\bibitem{DBLP:journals/ppl/HellerHLMSS07}
Steve Heller, Maurice Herlihy, Victor Luchangco, Mark Moir,
William N. Scherer III, and Nir Shavit. 2007. A {L}azy {C}oncurrent {L}ist-{B}ased {S}et {A}lgorithm. {\em Parallel Processing Letters} 17, 4 (2007), 411-424.


\bibitem{Herlihy90}
Maurice Herlihy and Jeannette M. Wing. 1990. Linearizability: {A} {C}orrectness {C}ondition for {C}oncurrent {O}bjects. {\em {ACM} Trans. Program. Lang. Syst.} 12, 3 (1990), 463-492.



\bibitem{Victor08}
Viktor Vafeiadis. 2008. {\em Modular {F}ine-{G}rained {C}oncurrency 
{V}erification.} Technical Report UCAM-CL-TR-726. University of Cambridge, Computer Laboratory.


\bibitem{Victor10}
Viktor Vafeiadis. 2010. Automatically {P}roving {L}inearizability.
In {\em CAV 2010, LNCS vol. 6174}. Springer. 450-464.



\bibitem{Feng12}
Hongjin Liang, Xinyu Feng, and Ming Fu. 2012. A {R}ely-{G}uarantee-{B}ased {S}imulation for {V}erifying {C}oncurrent {P}rogram {T}ransformations. In {\em POPL 2012}, ACM, 455-468.



\bibitem{Feng13}
Hongjin Liang and Xinyu Feng. 2013. Modular {V}erification of {L}inearizability with {N}on-{F}ixed {L}inearization Points. In {\em PLDI 2013},
ACM, 459-470.


\bibitem{Liu13}
Yang Liu, Wei Chen, Yanhong A. Liu, Jun Sun, Shao Jie Zhang and
Jin Song Dong. 2013. Verifying {L}inearizability via {O}ptimized {R}efinement {C}hecking. {\em {IEEE} Trans. Software Eng.} 39, 7 (2013), 1018-1039.



\bibitem{Alur10}
Pavol Cern{\'{y}}, Arjun Radhakrishna, Damien Zufferey, Swarat Chaudhuri, and
Rajeev Alur. 2010. Model {C}hecking of {L}inearizability of {C}oncurrent {L}ist {I}mplementations. In {\em CAV 2010 (LNCS vol.6174)}. Springer, 465-479.


\bibitem{Glabbeek96}
Rob J. van Glabbeek and W. P. Weijland. 1996. Branching {T}ime and {A}bstraction in {B}isimulation {S}emantics. {em J. ACM} 43, 3 (1996), 555-600.


\bibitem{Namjoshi97}
Kedar S. Namjoshi. 1997. A {S}imple {C}haracterization of {S}tuttering {B}isimulation. In {\em FSTTCS, LNCS 1346}, 284-296. 


\bibitem{Colvin06}
Robert Colvin, Lindsay Groves, Victor Luchangco, and Mark Moir. 2006. Formal {V}erification of a {L}azy {C}oncurrent {L}ist-{B}ased {S}et {A}lgorithm. 
In {\em CAV 2006, LNCS vol. 4144}. Springer, 475-488.


\bibitem{Derrick11}
John Derrick, Gerhard Schellhorn, and Heike Wehrheim. 2011. Verifying {L}inearisability with {P}otential {L}inearisation {P}oints. In {\em FM 2011 LNCS vol. 6664}. Springer, 323-337.


\bibitem{Schellhorn12}
Gerhard Schellhorn, Heike Wehrheim, and John Derrick. 2012.
How to {P}rove {A}lgorithms {L}inearisable. In {\em CAV 2012, LNCS vol.7358}.
Springer. 243-259.


\bibitem{spin09}
Martin T. Vechev, Eran Yahav, and Greta Yorsh. 2009. Experience with {M}odel {C}hecking {L}inearizability. In {\em SPIN 2009, LNCS vol. 5578}. Springer. 261-278.


\bibitem{pldi10}
Sebastian Burckhardt, Chris Dern, Madanlal Musuvathi, and Roy Tan. 2010.
Line-up: {A} {C}omplete and {A}utomatic {L}inearizability {C}hecker.
In {\em PLDI 2010}, ACM, 330-340.



\bibitem{Groote90}
Jan Friso Groote and Frits W. Vaandrager. 1990. An {E}fficient {A}lgorithm for {B}ranching {B}isimulation and {S}tuttering {E}quivalence. In {\em ICALP 1990 LNCS vol. 443}. Springer. 626-638.


\bibitem{arXiv17}
Xiaoxiao Yang, Joost{-}Pieter Katoen, Huimin Lin, and Hao Wu. 2017.
Verifying Concurrent Stacks by Divergence-Sensitive Bisimulation.
{\em CoRR} abs/1701.06104 (2017).



\bibitem{arXiv16}
Xiaoxiao Yang, Joost{-}Pieter Katoen, Huimin Lin, and Hao Wu. 2016.
Proving Linearizability via Branching Bisimulation. {\em CoRR} abs/1609.07546 (2016).


\bibitem{DBLP:conf/popl/DoddsHK15}
Mike Dodds, Andreas Haas, and Christoph M. Kirsch. 2015. A Scalable, Correct Time-Stamped Stack. In {\em POPL 2015}. 233-246.


\bibitem{DBLP:conf/esop/KhyzhaDGP17}
Artem Khyzha, Mike Dodds, Alexey Gotsman, and Matthew J. Parkinson. 2017.
Proving Linearizability Using Partial Orders. In {\em ESOP}. 639-667.


\bibitem{DBLP:conf/tacas/GrooteW16}
Jan Friso Groote and Anton Wijs. 2016. An O(m{\textbackslash}log n) Algorithm for Stuttering Equivalence and Branching Bisimulation. In {\em TACAS}. 607-624.

\bibitem{DBLP:conf/ifip/Jones83}
Cliff B. Jones. 1983. Specification and Design of (Parallel) Programs.
In {\em {IFIP} Congress}. 321-332.



\bibitem{DBLP:journals/tcs/AbadiL91}
Mart{\'{\i}}n Abadi and Leslie Lamport. 1991. 
The Existence of Refinement Mappings. {\em Theor. Comput. Sci.}. 82, 2, 253-284. (1991).


\bibitem{cegar}
E. Clark, O. Grumberg, etc.
Counterexample-Guided Abstraction Refinement.
In: CAV, LNCS 1855, pages 154-169. 2000.

\bibitem{book08}
C. Baier and J.-P. Katoen.
Principles of Model Checking. The MIT Press. 2008.

\bibitem{DBLP:conf/icalp/BaetenG87}
Jos C. M. Baeten and Rob J. van Glabbeek. 1987. Another Look at Abstraction in Process Algebra (Extended Abstract). {\em ICALP87}. 84-94.





%
%
%
%
%
%
%
%
%
%
%
%
%
%
%
%
%
%
%
%
%
%
%
%
%
%
%
%
%
%
%
%
%
%
%
%
%


\end{thebibliography}


\end{document}